\documentclass[11pt]{article}
\usepackage[T1]{fontenc}
\usepackage[utf8]{inputenc}
\usepackage{mathpazo}
\usepackage[margin=1in]{geometry}
\usepackage{amsmath,amssymb,amsthm,mathtools}
\usepackage{microtype}
\usepackage{graphicx}
\usepackage{booktabs,array}
\usepackage[most]{tcolorbox}
\usepackage[numbers,sort&compress]{natbib}
\usepackage[colorlinks=true,linkcolor=blue,citecolor=blue,urlcolor=blue]{hyperref}
\allowdisplaybreaks[1]
\numberwithin{equation}{section}
\newtheorem{theorem}{Theorem}[section]
\newtheorem{lemma}[theorem]{Lemma}

\newtheorem{corollary}[theorem]{Corollary}
\theoremstyle{definition}
\newtheorem{definition}[theorem]{Definition}

\theoremstyle{remark}
\newtheorem{remark}[theorem]{Remark}

\newtcolorbox{questionbox}{enhanced, colback=blue!4, colframe=blue!50!black,
  boxrule=0.6pt, arc=2pt, left=8pt, right=8pt, top=6pt, bottom=6pt}

\providecommand{\Lin}{\mathcal{L}}            
\providecommand{\Dens}{\mathcal{D}}           
\providecommand{\Tr}{\operatorname{Tr}}

\newcommand{\R}{\mathbb R}

\newcommand{\ket}[1]{\lvert #1\rangle}
\newcommand{\bra}[1]{\langle #1\rvert}
\newcommand{\braket}[2]{\langle #1\mid #2\rangle}
\newcommand{\id}{\operatorname{id}}
\newcommand{\ii}{\mathrm i}

\title{Stinespring's Dilation, Universal-NOT, and \\Limitations of Quantum Uncoupled Learning}
\author{
 Emmanouil-Vasileios Vlatakis-Gkaragkounis\thanks{Department of Computer Sciences, University of Wisconsin-Madison, email: \texttt{vlatakis@wisc.edu}} \and 
Pucheng Xiong\thanks{Department of Computer Sciences, University of Wisconsin-Madison, email: \texttt{pxiong79@wisc.edu}}}
\date{}
\begin{document}
\maketitle

\begin{abstract}
Uncoupled learning, in which players know only their own payoff, are a
standard model of decentralized learning toward Nash equilibrium. We study
their quantum counterpart, where strategies are quantum states and, in
each round, a player receives a physical copy of their opponent's current
strategy and updates their own state by a quantum channel. This adds to
uncoupledness a constraint with no classical analogue: every update must be
physically realizable. In fact, this constraint is decisive in
\emph{conjugation games}, in which Bob has a strictly dominant pure strategy
and Alice's unique best response is its transpose. In this work, we show some limitations of uncoupled (learning) dynamics in finding the Nash equilibrium for conjugation games. First, we prove that no autonomous, memoryless, uncoupled
dynamics converge to Nash equilibrium in a family of four single-qubit
conjugation games. Specifically, any memoryless channel that keeps all four
equilibria fixed will leave Alice's strategy frozen. This result also extends to
$4^n$ games on $n$-qubit registers. Second, we show that this barrier
disappears with a small classical memory: for any finite family of $L$ pure
targets, a guess-and-test rule with $O(\log L)$ classical bits
converges to equilibrium, with an explicit bound on its convergence rate. Finally,
we prove that no finite memory, classical or quantum, suffices for the
continuum of all pure-qubit targets. Our proof of the impossibility for memoryless channels relies on Stinespring's
dilation theorem, while the impossibility for finite-memory channels relies on the optimal universal-NOT (in)fidelity bound by
Bu\v{z}ek, Hillery, and Werner.
\end{abstract}
\newpage
{ \tableofcontents}
\newpage
\newpage

\section{Introduction}
\label{sec:intro}

Nash equilibrium is the central solution concept of non-cooperative game
theory, yet its existence is guaranteed by a fixed-point argument that says
nothing about how players might actually arrive at one. A long line of work
therefore asks whether Nash equilibrium can be reached by \emph{adaptive}
behavior: players interact repeatedly, each adjusts her strategy according
to a simple rule, and one asks whether the resulting dynamics converge to
equilibrium. Hart and Mas-Colell~\cite{HartMasColell2003} isolated the
informational constraint that makes this question both natural and
difficult. A dynamics is \emph{uncoupled} if each player's update rule
depends on her own payoff function and on what she observes of the play,
but not on the payoff functions of the other players. This reflects the
reality of decentralized learning, in which no player knows what her
opponents are trying to achieve.

\paragraph{Uncoupled dynamics in classical games.}
Uncoupledness is already a severe restriction for classical games. Hart
and Mas-Colell showed that no deterministic, stationary, continuous-time
uncoupled dynamics converges to Nash equilibrium in every game, even when
the equilibrium is unique~\cite{HartMasColell2003}. Randomization and
memory partially restore the picture. In discrete time, uncoupled dynamics
that react only to the current period of play (1-recall) cannot guarantee
almost-sure convergence to pure equilibria in general, whereas recalling the
last two periods suffices, and finite-memory procedures based on random
experimentation can bring play close to equilibrium in every finite
game~\cite{HartMasColell2006}. Regret testing~\cite{FosterYoung2006} and its
analysis by Germano and Lugosi~\cite{GermanoLugosi2007}, as well as learning
by trial and error~\cite{Young2009}, achieve similar guarantees even when a
player does not observe her opponents' actions at all. 

On the positive side, Shamma and coauthors showed that a little internal
state goes a long way. Uncoupled fictitious play and gradient play with
derivative action can converge where their static counterparts do
not~\cite{ShammaArslan2005}, and higher-order gradient dynamics converge
locally to any given isolated completely mixed equilibrium, although no
single such dynamics works in every game~\cite{ToonsiShamma2023}. In
structured classes, uncoupled rules reach pure equilibria: joint strategy
fictitious play with inertia in potential
games~\cite{MardenArslanShamma2009}, and completely uncoupled payoff-based
rules in weakly acyclic games~\cite{MardenYoungArslanShamma2009}.

Babichenko~\cite{Babichenko2012} showed that such \emph{completely} uncoupled dynamics
cannot lead to pure equilibria in almost all games that have them, although
they can lead to approximate equilibria. Hart and Mansour~\cite{HartMansour2010} quantified the cost of uncoupledness through
communication complexity, and Milionis, Papadimitriou, Piliouras, and
Spendlove~\cite{MilionisPapadimitriouPiliourasSpendlove2023} used Conley
index theory to exhibit games in which every game dynamics fails to converge
to Nash equilibrium from some starting points.

A common thread runs through this literature. The obstructions are
informational, since a player does not know what her opponents want, and
they are overcome, when they can be, by memory and randomness that let the
players search for an equilibrium. In all of these models, however, the
play itself is classical. A player who observes her opponent's action can
always compute her own best response to it, so the difficulty lies in
\emph{coordinating} on an equilibrium, never in \emph{reacting} to what has
been observed.

\paragraph{Quantum games and learning.}
In a quantum game, each player's strategy is a quantum state (a density
matrix), and payoffs are expectations of observables on the joint state.
This framework, which goes back to Meyer~\cite{Meyer1999} and to Eisert,
Wilkens, and Lewenstein~\cite{EisertWilkensLewenstein1999}, has been placed
on general foundations by Gutoski and Watrous~\cite{GutoskiWatrous2007} and
Zhang~\cite{Zhang2012}. As for classical games, computing an approximate
Nash equilibrium of a general quantum game is
PPAD-complete~\cite{BostanciWatrous2022}. Learning dynamics for quantum games
have recently attracted considerable attention. Matrix multiplicative
weights computes approximate equilibria of quantum zero-sum
games~\cite{JainWatrous2009}, and optimistic variants achieve a quadratic
speedup~\cite{VasconcelosEtAl2025}. The continuous-time quantum replicator
dynamics exhibit conservation laws and Poincar\'e recurrence in zero-sum
games~\cite{JainPiliourasSim2022}, and analogous dynamics have been studied
for quantum common-interest games~\cite{LinPiliourasSimVarvitsiotis2025} and
with payoff-based feedback~\cite{LotidisEtAl2023}.

These learning rules are classical algorithms that \emph{output} density
matrices. Each player maintains a classical description of her strategy and
updates it from classical feedback, such as the gradient operator
$\Tr_B[R_A(I \otimes \sigma)]$, which requires a classical description of
the opponent's state $\sigma$, or a scalar payoff observation. If the
players are themselves quantum devices that interact by exchanging quantum
systems, this is no longer the right model. A player then receives the
opponent's strategy as a \emph{physical system} rather than as a matrix,
and whatever she does with it must be a physical operation, namely a
quantum channel, possibly acting together with a persistent internal
memory~\cite{KretschmannWerner2005,ChiribellaDArianoPerinotti2009}.

\paragraph{Physical realizability as a new constraint.}
This model adds to uncoupledness a constraint with no classical
counterpart: each player's update must be \emph{physically realizable}.
Quantum mechanics restricts which transformations of unknown states can be
carried out. Unknown states cannot be cloned~\cite{WoottersZurek1982,Dieks1982},
a fixed device cannot be programmed to perform arbitrary unitaries with a
finite program~\cite{NielsenChuang1997}, and neither the universal-NOT nor
the transpose of an unknown state can be implemented
exactly~\cite{BuzekHilleryWerner1999,GisinPopescu1999,Buscemi2003}. The
amount of classical information that can be extracted from finitely many
copies of an unknown state is likewise limited~\cite{MassarPopescu1995}.

The simplest game in which these limits become strategic is what we call a
\emph{conjugation game}. Bob's payoff rewards him for playing a fixed pure
state $P_\psi$, which is therefore his strictly dominant strategy. Alice's
payoff is $\Tr[\rho\,\sigma^\top]$, so her unique best response to Bob's
dominant strategy is its transpose $P_\psi^\top = P_{\overline{\psi}}$. The
classical analogue of this game is trivial: Bob plays his dominant action,
Alice best-responds to the action she has just observed, and memoryless
best-response dynamics reach the equilibrium within two rounds. In the
quantum game, by contrast, Alice's best-response map $\sigma \mapsto
\sigma^\top$ is transposition, which is not a quantum channel. The
equilibrium is easy to describe, but the most natural way of reaching it is
physically impossible. This leads to the question studied in this paper.

\begin{questionbox}
\textbf{Question:} Consider uncoupled quantum dynamics, in which each player knows only her own
payoff observable and, in every round, receives a physical copy of her
opponent's current quantum strategy and updates her own state by a quantum
channel. \emph{Can such dynamics converge to Nash equilibrium, and how much
memory do the players need for this?}
\end{questionbox}

\subsection{Our contributions}
\label{sec:intro-contrib}

We answer this question for families of conjugation games, and the answer
depends sharply on the memory available to Alice. Without memory, the
dynamics cannot converge even in a family of four single-qubit games. A
few classical bits suffice for any finite family. No finite memory,
classical or quantum, suffices for the continuum of all pure targets.
Table~\ref{tab:intro-summary} summarizes the three results. In each case
Bob's rule is unrestricted, or as simple as possible, so that the
difficulty lies entirely with Alice.

\begin{table}[t]
\centering
\small
\begin{tabular}{@{}>{\raggedright\arraybackslash}p{3.0cm}
                >{\raggedright\arraybackslash}p{3.9cm}
                >{\raggedright\arraybackslash}p{4.2cm}
                >{\raggedright\arraybackslash}p{3.4cm}@{}}
\toprule
Alice's memory & Family of targets & Outcome & Result \\
\midrule
None & four qubit-states ($4^n$ states for $n$ qubits)
  & no convergence; Alice's strategy is frozen
  & Theorems~\ref{thm:four-game} and~\ref{thm:nqubit} \\
\addlinespace
$O(\log L)$ classical bits & any $L$ distinct pure states in $\mathbb{C}^d$
  & convergence, with explicit mixing time
  & Theorem~\ref{thm:fc} \\
\addlinespace
Any finite dimension, classical or quantum & all pure qubit-states
  & no convergence for some target
  & Theorem~\ref{thm:fm-barrier} and Corollary~\ref{cor:fm-games} \\
\bottomrule
\end{tabular}
\caption{Summary of results for conjugation games. All rules are
autonomous and uncoupled, and Alice receives a fixed number of copies of
Bob's strategy in each round.}
\label{tab:intro-summary}
\end{table}

\paragraph{Memoryless dynamics cannot converge to Nash equilibrium.}
Our first result (Theorem~\ref{thm:four-game}) exhibits four
single-qubit conjugation games, with targets $\ket{0}$, $\ket{1}$,
$\ket{+}$, and $\ket{{+i}}$, that share Alice's payoff observable. We show
that no autonomous, memoryless, uncoupled dynamics converges to Nash
equilibrium from every initial profile in all four games, whatever rules
Bob uses. The conclusion is in fact much stronger than non-convergence: any
memoryless channel for Alice that merely leaves the four equilibria fixed
must be the partial trace over Bob's system, so Alice's strategy never
moves at all. The construction extends to $4^n$ games on $n$-qubit
registers, even when the players' channels act collectively on all qubits
(Theorem~\ref{thm:nqubit}). The contrast with the classical setting is
stark, since the classical counterpart of these games is solved by
memoryless best-response dynamics.

The proof is a rigidity argument in the spirit of the no-cloning and
no-programming theorems. Convergence forces each equilibrium to be a fixed
point of Alice's channel, and each fixed-point condition asks for a
\emph{pure} output. By Stinespring's dilation theorem, a channel with a
pure output is uncorrelated with its environment, so everything it takes
from Bob's copy must be deposited, intact, in the environment. Because
channels preserve inner products at the level of the dilation, the
environment states must reproduce the overlaps of Bob's states, so the
environment holds an isometric copy of Bob's input and Alice's register
learns nothing. The complex target $\ket{{+i}}$ is essential: it makes the
four constraints span the whole input space and thereby determine the
channel completely. Indeed, if all targets are real, then
$P_{\overline{\psi}} = P_\psi$ and Alice can simply copy Bob's state. The
obstruction is therefore exactly complex conjugation.

\paragraph{Finite classical memory suffices for finite families.}
Our second result (Theorem~\ref{thm:fc}) shows that this barrier disappears
as soon as Alice has a small classical memory. For any finite family of $L$
distinct pure targets in any dimension, we give an autonomous uncoupled
rule in which Alice stores a label from $\{1,\dots,L\}$, and we prove that
the joint memory--strategy state converges to equilibrium with mixing time
at most $\lceil (2(L-1) + 8\ln(1/\varepsilon))/\delta \rceil$, where
$\delta$ is the smallest probability with which a wrong target is detected.
Two classical bits suffice for the four-game family, and $2n$ bits suffice
for its $n$-qubit extension (Corollary~\ref{cor:fc-families}).

The key idea is to replace an impossible task by a possible one. Alice
cannot \emph{transform} Bob's unknown state into its conjugate, but she can
\emph{identify} which of finitely many known targets he is playing and then
prepare the corresponding best response from its classical description.
Her label is a current guess, which she checks in each round by a two-outcome
measurement on Bob's fresh copy. A correct guess is never rejected, because
a pure target passes its own test with certainty, and a wrong guess is
rejected with probability at least $\delta$, so the label settles on the
truth after finitely many rounds. The rule escapes the rigidity argument
because the classical label makes the equilibrium states of different games
orthogonal, and orthogonal inputs place no constraint on the overlaps that
drove the first result. In this sense, classical memory converts a quantum
distinction (between nonorthogonal states) into a classical one (between
orthogonal labels).

\paragraph{No finite memory suffices for the continuum.}
Our third result (Theorem~\ref{thm:fm-barrier}) shows that the success of
the second cannot be extended to the family of \emph{all} pure qubit
targets. For any finite-dimensional persistent register, classical or
quantum, any fixed number of copies of Bob's state per round, and any update
channel and initial state, there is a pure state $P$ such that Alice's
strategy does not converge to her best response $P^\top$, even though Bob
plays $P$ in every round. As a consequence, if Bob's rule retains his
strictly dominant strategy once he plays it, then no autonomous uncoupled
dynamics in which Alice has finite memory converges to Nash equilibrium in
every game of this continuum (Corollary~\ref{cor:fm-games}). The retention
assumption cannot be dropped, since otherwise Bob could signal Alice's
answer by playing $P^\top$ in the first round (Remark~\ref{rem:fm-retention}).
This again contrasts with the classical picture, where two periods of recall
suffice for convergence to pure equilibria~\cite{HartMasColell2006}: here
every game has a unique equilibrium in pure states and Bob has a strictly
dominant strategy, yet no finite memory suffices.

The proof plays two rates against each other. On the one hand, for a fixed
target, Alice's register evolves by powers of a single channel on a
finite-dimensional space. Such channels are power-bounded, so their dynamics split into a recurrent
part, which keeps returning near its starting point and therefore cannot
contribute to an error that tends to zero, and a transient part, which
decays geometrically. This splitting is the finite-dimensional case of the
Jacobs--de~Leeuw--Glicksberg (JdLG) decomposition of operator
semigroups~\cite{Jacobs1956,Jacobs1957,deLeeuwGlicksberg1961} (see
also~\cite[Ch.~16]{EisnerFarkasHaaseNagel2015}). For quantum channels, the
recurrent part is known as the attractor, or peripheral,
subspace~\cite{AmatoFacchiKonderak2023}. It can be considered as a linear,
quantitative counterpart of Conley's decomposition of dynamical
systems~\cite{Conley1978}, and it implies that a finite-memory learner
converges exponentially fast or not at all. The Baire category theorem then
gives a uniform exponential rate on an open patch of targets. By coarsely
estimating the unknown state and rotating it into this patch, we convert the
learner into a device that transposes \emph{every} pure state with error
exponentially small in the number of copies used.

On the other hand,
transposition differs from the universal-NOT only by a fixed unitary, and
the optimal universal-NOT of Bu\v{z}ek, Hillery, and
Werner~\cite{BuzekHilleryWerner1999} shows that any such device has average
error at least $1/(N+2)$ on $N$ copies. A finite-memory learner that always
succeeded would thus be faster than quantum mechanics permits.

\subsection{Further related work, organization, and AI disclosure}
\paragraph{Further related work.}
Our first result belongs to a family of structure theorems for operations
that leave a set of states undisturbed~\cite{KoashiImoto2002,KretschmannSchlingemannWerner2008},
and our use of the optimal universal-NOT relates the dynamics to optimal
approximations of the transpose map~\cite{Buscemi2003} and to the extra
information carried by antiparallel spins~\cite{GisinPopescu1999}. The recurrent/transient splitting behind our third result is the
Jacobs--de~Leeuw--Glicksberg
decomposition~\cite{Jacobs1956,Jacobs1957,deLeeuwGlicksberg1961}, which
also underlies the asymptotic theory of quantum channels and quantum Markov
semigroups~\cite{AmatoFacchiKonderak2023,AmatoFacchiKonderak2025}. We use
it only through one consequence: for a finite-memory learner, an error that
tends to zero must do so exponentially fast. The
study of learning under memory constraints goes back to Hellman and
Cover~\cite{HellmanCover1970}, who developed the theory of optimal
finite-memory procedures for hypothesis testing. The rule of our second
result is a finite automaton driven by independent quantum observations in
the same spirit, but it attains zero asymptotic error because a pure target
passes its own test with certainty. Our barriers concern exact convergence
to equilibrium; whether finite memory suffices for convergence to
\emph{approximate} equilibria over the continuum of targets is not
addressed here.

\paragraph{Organization.}
Section~\ref{sec:prelim} introduces quantum channels, quantum games,
conjugation games, and the model of uncoupled dynamics with memory, and it
collects the consequences of Stinespring's theorem that we use. Section \ref{sec:warmup} shows that the classical impossibility of Hart and Mas-Collel \cite{HartMasColell2003} can be "embedded" into the quantum setting. Section~\ref{sec:memoryless-impossibilty} proves the impossibility for
memoryless dynamics and its $n$-qubit extension.
Section~\ref{sec:finite-classical} constructs the finite-memory rule and
analyzes its mixing time. Section~\ref{sec:barrier} proves the barrier for
finite quantum memory. Section \ref{sec:conclusion} is the conclusion section and it points out two further directions based on the main techniques used in this paper.

\paragraph{Disclosure of AI use.}
In preparing this paper, the author used generative AI tools (ChatGPT
6-Astra, Fable 5.2, and Opus 5.5), following the author's instructions to help prepare
the manuscript. These tasks included elaborating
on and polishing the writing and exposition, auditing proofs, and generating figures. The mathematical
ideas are primarily the author's own. The AI tools were used to elaborate on
the author's ideas, specifically on the use of Stinespring's dilation theorem
and the analysis of the convergence of power-bounded operators, and to work
out the details based on the author's original arguments. The authors take full responsibility for all results.

\section{Preliminaries}
\label{sec:prelim}
This section fixes notation, introduces the conjugation games used
throughout the paper, defines the model of uncoupled quantum dynamics with
memory, and collects the consequences of Stinespring's dilation theorem
on which our impossibility results rest. All Hilbert spaces are finite
dimensional. Transposes $X^\top$ and complex conjugates $\overline{X}$,
$\overline{\psi}$ are taken in a fixed orthonormal basis.
 
\paragraph{Quantum states.}
The state of a system with Hilbert space $H$ is a density matrix
$\rho \in \Dens(H) = \{\rho \in \Lin(H) : \rho \succeq 0,\ \Tr[\rho] = 1\}$.
Pure states are rank-one projectors $P_\psi = \ket{\psi}\bra{\psi}$. For a
pure state $P$, $\Tr[P\rho] \le 1$ with equality if and only if $\rho = P$.
A qubit-state is $P(r) = \frac12(I + r_1 X + r_2 Y + r_3 Z)$ with
$\|r\|_2 \le 1$, and it is pure exactly when $\|r\|_2 = 1$. Independent
systems are combined by the tensor product. The partial trace, defined by
$\Tr_B[Y \otimes Z] = \Tr[Z]\,Y$, gives the state of $A$ alone.
 
\paragraph{Quantum channels.}
Every physical operation that can be applied to an unknown input state is
described by a quantum channel.
 
\begin{definition}[Quantum channel]
A linear map $\Phi : \Lin(H_{\mathrm{in}}) \to \Lin(H_{\mathrm{out}})$ is a
\emph{quantum channel} if it is trace preserving and \emph{completely
positive}: $\Phi \otimes \id_{\Lin(K)}$ maps positive operators to positive
operators for every finite-dimensional ancilla space $K$.
\end{definition}
 
Unitary evolution, appending fresh ancillas, measurements, private
randomness, discarding subsystems, and all compositions of these
operations are channels. A linear map on
$\Lin(\mathbb{C}^d)$ is completely positive if and only if its Choi matrix
$\sum_{a,b}\ket{a}\bra{b}\otimes\Phi(\ket{a}\bra{b})$ is positive
semidefinite~\cite{Choi1975}. The transpose map $T(X) = X^\top$ preserves
density matrices, and $P_\psi^\top = P_{\overline{\psi}}$. However, its
Choi matrix is the swap operator, which has eigenvalue $-1$, so $T$ is
\emph{not} a channel: no physical device takes an unknown state $\rho$ and
returns $\rho^\top$. On a qubit, $Y^\top = -Y$ while $X$ and $Z$ are
symmetric, so $T$ reflects the Bloch sphere through the $x$--$z$ plane. It
fixes real states such as $\ket{0}$, $\ket{1}$, $\ket{+}$, and maps
$\ket{{+i}} = \tfrac{1}{\sqrt{2}}(\ket{0}+i\ket{1})$ to $\ket{{-i}}$.
 
\paragraph{Quantum games and Nash equilibrium.}
Alice and Bob choose strategies $\rho \in \Dens(H_A)$ and
$\sigma \in \Dens(H_B)$. A game $G = (R_A, R_B)$ is given by Hermitian
payoff observables on $H_A \otimes H_B$, with
$u_A(\rho,\sigma) = \Tr[R_A(\rho\otimes\sigma)]$ and
$u_B(\rho,\sigma) = \Tr[R_B(\rho\otimes\sigma)]$. When $R_A$ and $R_B$ are
diagonal in a product basis, this is a classical bimatrix game in mixed
strategies; entangled observables give genuinely quantum games. A profile
$(\rho^*,\sigma^*)$ is a \emph{Nash equilibrium} if
$u_A(\rho,\sigma^*) \le u_A(\rho^*,\sigma^*)$ and
$u_B(\rho^*,\sigma) \le u_B(\rho^*,\sigma^*)$ for all $\rho$ and $\sigma$.
A state is a \emph{unique dominant strategy} for Bob if it is his unique
maximizer against every strategy of Alice.
 
We now introduce the \emph{conjugation games}, which underlie all of our
results. Their key ingredient is Alice's payoff observable, which is built
from the maximally entangled state. For $H_A = H_B = \mathbb{C}^d$, let
\[
  \ket{\Omega_d} = \frac{1}{\sqrt{d}}\sum_{a}\ket{a}\otimes\ket{a}.
\]
A direct computation gives
\begin{equation}\label{eq:prelim-transpose-payoff}
  d\,\bra{\Omega_d}\rho\otimes\sigma\ket{\Omega_d}
  = \textstyle\sum_{a,b}\rho_{ab}\,\sigma_{ab} = \Tr[\rho\,\sigma^\top],
\end{equation}
so the observable $d\,\ket{\Omega_d}\bra{\Omega_d}$ rewards Alice for
matching the transpose of Bob's state.
\begin{lemma}[Conjugation game]\label{lem:prelim-conj-ne}
Let $R_A = d\,\ket{\Omega_d}\bra{\Omega_d}$ and $R_B = I \otimes P_\psi$ for a
pure state $P_\psi$. Then $P_\psi$ is Bob's unique dominant strategy,
Alice's unique best response to it is $P_\psi^\top = P_{\overline{\psi}}$,
and $(P_{\overline{\psi}}, P_\psi)$ is the unique Nash equilibrium.
\end{lemma}
 
\begin{proof}
Apply $\Tr[P\rho] \le 1$, with equality only at $\rho = P$, to
$u_B = \Tr[P_\psi\sigma]$ and to
$u_A(\rho, P_\psi) = \Tr[\rho P_\psi^\top]$, which follows
from~\eqref{eq:prelim-transpose-payoff}.
\end{proof}
 
Thus Alice's best-response map is the transposition of Bob's state, which
is precisely the operation that no channel can perform. We call $G = (R_A,
R_B)$ as in Lemma~\ref{lem:prelim-conj-ne} the \emph{conjugation game with
target} $\psi$.
 
\paragraph{Uncoupled, autonomous, and memoryless dynamics.}
The game is played repeatedly in rounds $t = 0,1,2,\dots$. Players never see classical
descriptions of each other's strategies; instead, in each round a player
receives a physical copy of the opponent's current state. Alice's
\emph{persistent register} is $S_A = H_A \otimes M_A$, where $H_A$ holds her
strategy and $M_A$ is a finite-dimensional \emph{memory}. Everything kept
for a later round counts as memory, including stored measurement outcomes,
classical labels, and clocks. Ancillas, randomness, and outcomes that are
used and discarded within a round do not count. This is the model of
quantum channels with memory of Kretschmann and
Werner~\cite{KretschmannWerner2005}: each use of the channel acts jointly
on a fresh input and on a memory system, which is passed on to the next
use. Here the fresh input is the copy of the opponent's state, and the
memory system is Alice's persistent register $S_A$.
 
\begin{definition}[Update rules]\label{def:prelim-rule}
An update rule for Alice consists of an initial state
$\omega_{A,0}\in\Dens(S_A)$ and channels
$\Phi_{A,t} : \Lin(S_A \otimes H_B) \to \Lin(S_A)$, with
$\omega_{A,t+1} = \Phi_{A,t}(\omega_{A,t}\otimes\sigma_t)$ and strategy
$\rho_t = \Tr_{M_A}[\omega_{A,t}]$. The rule is \emph{autonomous} if
$\Phi_{A,t} = \Phi_A$ for all $t$, and \emph{memoryless} if
$\dim M_A = 1$. Rules assigned to a family of games are
\emph{uncoupled}~\cite{HartMasColell2003} if Alice's channels and initial
memory are the same in any two games with the same $R_A$, and likewise
for Bob.
\end{definition}
 
An autonomous memoryless update rule therefore has neither a clock nor a
private record of past rounds. With channels $\Phi_A : \Lin(H_A\otimes H_B)\to\Lin(H_A)$ and
$\Psi_B : \Lin(H_A\otimes H_B)\to\Lin(H_B)$, the profile evolves by
\begin{equation}\label{eq:prelim-memoryless}
  \rho_{t+1} = \Phi_A(\rho_t\otimes\sigma_t), \qquad
  \sigma_{t+1} = \Psi_B(\rho_t\otimes\sigma_t).
\end{equation}
Since channels are continuous, any limit $(\rho^*,\sigma^*)$ of a trajectory
of~\eqref{eq:prelim-memoryless} is a fixed point:
$\Phi_A(\rho^*\otimes\sigma^*) = \rho^*$. Now consider a family of
conjugation games with common $R_A = d\,\ket{\Omega_d}\bra{\Omega_d}$.
Uncoupledness assigns Alice a single channel $\Phi_A$ for the whole
family. If the dynamics converge to equilibrium in every game, then
Lemma~\ref{lem:prelim-conj-ne} gives
\begin{equation}\label{eq:prelim-fixed-condition}
  \Phi_A(P_{\overline{\psi}}\otimes P_\psi) = P_{\overline{\psi}}
  \qquad \text{for every target } \psi \text{ in the family.}
\end{equation}
 
\begin{remark}\label{rem:prelim-not-trivial}
Condition~\eqref{eq:prelim-fixed-condition} does not require Alice to
transpose, because at equilibrium she already holds $P_{\overline{\psi}}$.
It is satisfied by $\Phi_A = \Tr_B$, which keeps Alice's state and never
moves it. Complex amplitudes matter here. If all targets are real, then
$P_{\overline{\psi}} = P_\psi$, and the rule in which Alice copies Bob's
register while Bob prepares $P_\psi$ reaches equilibrium within two
rounds.
\end{remark}
 
\paragraph{Stinespring's dilation theorem.}
Stinespring's theorem states that every channel arises in the same way:
the input is embedded reversibly into a larger system, and part of that
system, called the \emph{environment}, is then discarded. A channel thus
never destroys information; it only moves information into the
environment. This viewpoint is the main tool of
Section~\ref{sec:memoryless-impossibilty}.
 
\begin{theorem}[Stinespring~\cite{Stinespring1955}]
\label{thm:prelim-stinespring}
A linear map $\Phi : \Lin(H_{\mathrm{in}}) \to \Lin(H_{\mathrm{out}})$ is
a quantum channel if and only if there are a finite-dimensional Hilbert
space $E$ and an isometry
$V : H_{\mathrm{in}} \to H_{\mathrm{out}} \otimes E$ such that
\begin{equation}\label{eq:prelim-stinespring}
  \Phi(X) = \Tr_E\bigl[V X V^\dagger\bigr]
  \qquad \text{for all } X \in \Lin(H_{\mathrm{in}}).
\end{equation}
One may take
$\dim E \le \dim H_{\mathrm{in}} \cdot \dim H_{\mathrm{out}}$.
\end{theorem}
 
The pair $(V, E)$ is a \emph{Stinespring dilation} of $\Phi$. Choosing
an orthonormal basis $\{\ket{k}\}$ of $E$ and setting
$V = \sum_k K_k \otimes \ket{k}$ recovers the equivalent Kraus operator form
$\Phi(X) = \sum_k K_k X K_k^\dagger$ with
$\sum_k K_k^\dagger K_k = I$. For example, the partial trace
$\Tr_B : \Lin(H_A \otimes H_B) \to \Lin(H_A)$ has the dilation
$V = I_A \otimes W$ for any isometry $W : H_B \to E$, where Bob's system is
moved into the environment and then discarded.
 
For a pure input $P_\psi$, the dilated output $V\ket{\psi}$ is a pure
state of the output system and the environment, and $\Phi(P_\psi)$ is its
marginal on the output. This picture has two simple but important
consequences.

\begin{lemma}[Pure outputs and overlaps]\label{lem:prelim-pure-output}
Let $(V, E)$ be a Stinespring dilation of $\Phi$, and let
$\psi_1, \dots, \psi_m \in H_{\mathrm{in}}$ be unit vectors such that
each output $\Phi(P_{\psi_j}) = P_{\phi_j}$ is pure. Then
\begin{enumerate}
  \item there are unit vectors $e_j \in E$ with
        $V\ket{\psi_j} = \ket{\phi_j} \otimes \ket{e_j}$; and
  \item for all $j, k$,
        \begin{equation}\label{eq:prelim-overlap}
          \braket{\psi_j}{\psi_k}
          = \braket{\phi_j}{\phi_k}\, \braket{e_j}{e_k}.
        \end{equation}
\end{enumerate}
\end{lemma}
 
\begin{proof}
Write the Schmidt decomposition
$V\ket{\psi_j} = \sum_k \sqrt{\lambda_k}\,\ket{a_k}\otimes\ket{b_k}$ with
orthonormal families $\{a_k\}$, $\{b_k\}$ and $\lambda_k > 0$. Its
marginal on $H_{\mathrm{out}}$ is
$\sum_k \lambda_k\ket{a_k}\bra{a_k} = P_{\phi_j}$, which has rank one, so
there is a single term, and $\ket{a_1} = c\ket{\phi_j}$ with $|c| = 1$.
Setting $\ket{e_j} = c\ket{b_1}$ proves part~1. For part~2, an isometry
preserves inner products, so
$\braket{\psi_j}{\psi_k} = \bra{\psi_j} V^\dagger V \ket{\psi_k}
 = \braket{\phi_j}{\phi_k}\, \braket{e_j}{e_k}$.
\end{proof}
 
Part~1 says that when an output is pure, it is completely uncorrelated
with the environment: every part of the input that does not appear in
the output has gone, in its entirety, into the environment. Part~2 then
records how input overlaps are split between output and environment.
Since $|\braket{e_j}{e_k}| \le 1$, it implies
$|\braket{\psi_j}{\psi_k}| \le |\braket{\phi_j}{\phi_k}|$: a channel that
maps pure states to pure states cannot make them more distinguishable.
For instance, under~\eqref{eq:prelim-fixed-condition} the inputs are
$\overline{\psi}\otimes\psi$ and the outputs are $\overline{\psi}$, so
\eqref{eq:prelim-overlap} reads
$\overline{\braket{\psi}{\phi}}\,\braket{\psi}{\phi}
 = \overline{\braket{\psi}{\phi}}\,\braket{e_\psi}{e_\phi}$. Hence
$\braket{e_\psi}{e_\phi} = \braket{\psi}{\phi}$ whenever
$\braket{\psi}{\phi} \neq 0$. That is, the environment vectors reproduce the
geometry of Bob's states exactly. The next lemma turns such an equality
of overlaps into an isometry.
 
\begin{lemma}[Equal Gram matrices give an isometry]\label{lem:prelim-gram}
Let $\psi_1, \dots, \psi_m$ span a Hilbert space $H$, and let
$e_1, \dots, e_m$ be vectors in a Hilbert space $E$ with
$\braket{e_j}{e_k} = \braket{\psi_j}{\psi_k}$ for all $j, k$. Then there
is a unique linear isometry $W : H \to E$ with $W\ket{\psi_j} = \ket{e_j}$
for every $j$.
\end{lemma}
 
\begin{proof}
Define $W\bigl(\sum_j c_j \ket{\psi_j}\bigr) = \sum_j c_j \ket{e_j}$. For
any coefficients $c \in \mathbb{C}^m$,
\[
  \Bigl\| \sum_j c_j \ket{e_j} \Bigr\|^2
  = \sum_{j,k} \overline{c_j}\, c_k\, \braket{e_j}{e_k}
  = \sum_{j,k} \overline{c_j}\, c_k\, \braket{\psi_j}{\psi_k}
  = \Bigl\| \sum_j c_j \ket{\psi_j} \Bigr\|^2 .
\]
If $\sum_j c_j \ket{\psi_j} = 0$, the right-hand side vanishes, and so
$\sum_j c_j \ket{e_j} = 0$. Hence $W$ is well defined. It is linear by
construction, and the identity shows that it preserves norms. Uniqueness
holds because the $\psi_j$ span $H$.
\end{proof}

\section{Warm-Up: The Hart--Mas-Colell Construction on the Bloch Ball}
\label{sec:warmup}

Before turning to the physical model of Definition~\ref{def:prelim-rule},
we ask what uncoupledness \emph{alone} implies for quantum games. In this
section we therefore grant the players the most generous feedback that one
could imagine. Each player sees a complete classical description of every
player's current state, and she may update her own state by an arbitrary
smooth rule, whether or not this rule is a physical operation. We show
that the impossibility theorem of Hart and
Mas-Colell~\cite{HartMasColell2003} survives in this setting: there is a
family of three-qubit games with unique Nash equilibria on which no smooth
uncoupled dynamics are Nash-convergent (Theorem~\ref{thm:warmup}).

The section has three parts. Section~\ref{sec:warmup-hmc} recalls the
classical construction of Hart and Mas-Colell and isolates the three ideas
behind it. Section~\ref{sec:warmup-games} lifts the construction from the
simplex to the Bloch ball and explains how each idea transfers.
Section~\ref{sec:warmup-proof} proves the impossibility theorem. The
argument is classical throughout and uses no property of quantum channels.
Its purpose is to separate what quantum games inherit from the classical
theory from what is new: the obstruction of this section is
\emph{informational}, whereas the obstructions of
Sections~\ref{sec:memoryless-impossibilty} and~\ref{sec:barrier} are
\emph{physical}.

\subsection{The construction of Hart and Mas-Colell}\label{sec:warmup-hmc}

\paragraph{Uncoupled dynamics and Nash convergence.}
Hart and Mas-Colell~\cite{HartMasColell2003} consider continuous-time
dynamics on the mixed-strategy profiles $x = (x_1, \dots, x_n)$ of a game
with payoff functions $u_1, \dots, u_n$,
\begin{equation}\label{eq:warmup-hmc-dynamics}
  \dot x_i = F_i(x;\, u_i) \qquad (i = 1, \dots, n).
\end{equation}
The dynamics are \emph{uncoupled}: the update rule $F_i$ of player $i$ may
depend on the full profile $x$ and on her own payoff function $u_i$, but
not on the payoff functions of the other players. On a family of games
with unique Nash equilibria, the dynamics are \emph{Nash-convergent} if,
in every game of the family, the equilibrium is a rest point that attracts
every trajectory. As a regularity condition, $F$ is continuously
differentiable and the equilibrium is \emph{hyperbolically stable}: every
eigenvalue of the Jacobian at the equilibrium has negative real part. The
theorem of Hart and Mas-Colell states that there are families of games on
which no uncoupled dynamics are Nash-convergent. We recall their three-player game construction.

\paragraph{The three-player game.}
There are three players $i \in \mathbb{Z}_3$, and indices are read modulo
$3$, so that player $i+1$ is the \emph{successor} of player $i$ and player
$i-1$ is her \emph{predecessor}. Each player has two actions. The payoff of
player $i$ depends only on her own action and on that of her successor: she
receives $a_i$ if she plays her first action and her successor plays her
second, she receives $1$ if she plays her second action and her successor
plays her first, and she receives $0$ if the two actions agree. Each player
thus wants to \emph{mismatch} her successor. The base game $a_1 = a_2 = a_3 = 1$
is due to Jordan~\cite{Jordan1993}, and the family consists of all games
with $|a_i - 1| < \varepsilon$ for a small $\varepsilon > 0$. Let
$x_i \in [0,1]$ be the probability with which player $i$ plays her first
action. Her expected payoff is
\begin{equation}\label{eq:warmup-hmc-payoff}
  u_i(x) = a_ix_i\,(1 - x_{i+1}) + (1 - x_i) x_{i+1} .
\end{equation}
Three players on a cycle cannot all mismatch their successors, so the game
has no equilibrium in pure strategies. Player $i$ is willing to randomize
only if she is indifferent between her two actions; that is, only if
$a_i (1 - x_{i+1}) = x_{i+1}$. Every game of the family has a unique Nash
equilibrium, in which all three players are
indifferent:
\begin{equation}\label{eq:warmup-hmc-ne}
  x_i^* = \frac{a_{i-1}}{1 + a_{i-1}} \qquad (i \in \mathbb{Z}_3).
\end{equation}
In the base game, every player plays $\tfrac12$.

\paragraph{The argument for non-convergence.}
The proof that no uncoupled dynamics are Nash-convergent on this family
rests on following observations.
\begin{enumerate}
  \item \textbf{A hidden parameter.} The payoff
        function~\eqref{eq:warmup-hmc-payoff} of player $i$ involves only
        her own parameter $a_i$. The parameters $a_{i-1}$ and $a_{i+1}$ of
        the other two players are therefore invisible to an uncoupled rule
        of player $i$.
  \item \textbf{The equilibrium is controlled by the hidden parameter.}
        By~\eqref{eq:warmup-hmc-ne}, the equilibrium strategy of player
        $i$ depends on $a_{i-1}$ and not on $a_i$. At equilibrium, player
        $i$ plays exactly the strategy that makes her \emph{predecessor}
        indifferent, and this strategy is dictated by a payoff that she
        does not know.
  \item \textbf{The rest points force a traceless Jacobian.} Start from the base
        game and change $a_{i-1}$ alone. The payoff function of player $i$
        does not change, and neither do the equilibrium strategies of the
        other two players, which remain $\tfrac12$. The equilibrium
        strategy of player $i$, however, sweeps out a whole neighborhood
        of $\tfrac12$. Each of these profiles is a rest point in its own
        game, and by uncoupledness, player $i$ uses the same update rule
        in all of these games. Hence, in the base game, the update rule of
        player $i$ vanishes whenever the other two players are at
        $\tfrac12$ and she is anywhere near $\tfrac12$. Its derivative with
        respect to her own strategy is therefore zero at the equilibrium.
        This holds for every player, so the Jacobian at the equilibrium of
        the base game has zero diagonal and zero trace. The real parts of
        its eigenvalues then sum to zero, and they cannot all be negative.
\end{enumerate}
In short, a player cannot distinguish the game she is in from nearby games
whose equilibria ask her to stand somewhere else. She must therefore stay
put at all of these positions, and a system in which no player reacts to
her own displacement cannot be stable.

\paragraph{Bias coordinates.}
To transition from the classical construction to the quantum setting, we rewrite the game in variables that
are centered at the equilibrium of the base game. This is only a
reparametrization of the family of~\cite{HartMasColell2003}, where the
variables are $x_i$ and $a_i$.

In place of the probability $x_i$, we use
\[
  s_i = 2x_i - 1 \in [-1, 1].
\]
We call $s_i$ the \emph{bias} of player $i$: it equals $+1$ if she always
plays her first action, $-1$ if she always plays her second action, and
$0$ if she plays both with equal probability. In place of the payoff
weight $a_i$, we use
\[
  t_i = \frac{a_i - 1}{a_i + 1},
\]
which is $0$ in the base game $a_i = 1$ and is small throughout the
family. Substituting $x_i = (1 + s_i)/2$
into~\eqref{eq:warmup-hmc-payoff} gives
\begin{equation}\label{eq:warmup-hmc-bias}
  u_i = \frac{1 + a_i}{4}\,\Bigl[ s_i(t_i - s_{i+1})
        + 1 - t_i s_{i+1} \Bigr].
\end{equation}

Only the term $s_i\,(t_i - s_{i+1})$ matters for the decisions of player
$i$. The others do not involve her own strategy $s_i$, so neither of those terms changes the Nash
equilibria. For the purpose of finding equilibria, we may therefore take
the payoff of player $i$ to be
\begin{equation}\label{eq:warmup-bias-payoff}
  s_i(t_i - s_{i+1}) = s_it_i - s_is_{i+1}.
\end{equation}
This expression is easy to read. It is linear in $s_i$ with slope
$t_i - s_{i+1}$. If the slope is positive, her unique best response is
$s_i = +1$; if it is negative, her unique best response is $s_i = -1$; and
if it is zero, she is indifferent among all her strategies. In the base
game, $t_i = 0$, so she simply wants a bias of the sign opposite to that of
her successor, which is the mismatching game. A nonzero $t_i$ shifts the
point at which she is indifferent from $s_{i+1} = 0$ to $s_{i+1} = t_i$.

In the new variables, the equilibrium~\eqref{eq:warmup-hmc-ne} is
\begin{equation}\label{eq:warmup-hmc-ne-bias}
  s_i^* = t_{i-1} \qquad (i \in \mathbb{Z}_3),
\end{equation}
since $2a_{i-1}/(1 + a_{i-1}) - 1 = t_{i-1}$. Each player's successor
plays exactly the bias that makes her indifferent. In the base game the
equilibrium is $s^* = 0$.

\subsection{The quantum construction: three-qubit exchange games}
\label{sec:warmup-games}

We now replace each player's two actions by a qubit. The bias
$s_i \in [-1, 1]$ becomes a Bloch vector $r_i$ in the unit ball, the
parameter $t_i$ becomes a vector $\theta_i \in \R^3$, and the product
$s_i s_{i+1}$ becomes an inner product $r_i \cdot r_{i+1}$, which is the
expectation of a two-qubit observable.

\paragraph{Three-player qubit games.}
The players are again indexed by $i \in \mathbb{Z}_3$. Each player controls
one qubit and chooses a strategy $\rho_i \in \Dens(\mathbb{C}^2)$. A game
$G = (R_1, R_2, R_3)$ is given by Hermitian payoff observables on
$(\mathbb{C}^2)^{\otimes 3}$, with payoffs
$u_i(\rho_1,\rho_2,\rho_3) = \Tr[R_i(\rho_1\otimes\rho_2\otimes\rho_3)]$.
As in Section~\ref{sec:prelim}, a profile is a Nash equilibrium if no
player can increase her payoff by changing her own strategy.

\paragraph{Bloch coordinates.}
Write $\vec\sigma = (X, Y, Z)$ for the Pauli matrices and, for
$v = (v_x, v_y, v_z) \in \R^3$, let
$v\cdot\vec\sigma = v_x X + v_y Y + v_z Z$, so that
$P(v) = \tfrac12(I + v\cdot\vec\sigma)$. Throughout this section,
subscripts on $r$ and $\theta$ index players and not coordinates. The map
$v \mapsto P(v)$ is an affine bijection from the Bloch ball
$\mathbb{B} = \{v \in \R^3 : \|v\|_2 \le 1\}$ onto $\Dens(\mathbb{C}^2)$,
and it sends the boundary of the ball, the Bloch sphere, onto the pure
states. We therefore identify a strategy profile with a triple
$r = (r_1, r_2, r_3) \in \mathbb{B}^3$ of Bloch vectors, where
$\rho_i = P(r_i)$. Let
\[
  \Gamma = X\otimes X + Y\otimes Y + Z\otimes Z
\]
be the two-qubit \emph{exchange operator}. The Pauli matrices are
traceless, and for $A, B \in \{X, Y, Z\}$ the trace $\Tr[AB]$ equals $2$
if $A = B$ and $0$ otherwise. Thus, for all $v \in \R^3$ and
$r, r' \in \mathbb{B}$,
\begin{equation}\label{eq:warmup-bloch}
  \Tr\bigl[(v\cdot\vec\sigma)\,P(r)\bigr] = v\cdot r,
  \qquad
  \Tr\bigl[\Gamma\bigl(P(r)\otimes P(r')\bigr)\bigr] = r\cdot r' .
\end{equation}

\paragraph{Construction of the games.}
Fix $0 < \delta < \tfrac14$. A game of the family is specified by a
parameter $\theta = (\theta_1,\theta_2,\theta_3)$, where each
$\theta_i = (\theta_{i,x}, \theta_{i,y}, \theta_{i,z}) \in \R^3$ satisfies
$\|\theta_i\|_2 < \delta$. Recall that
$\theta_i\cdot\vec\sigma = \theta_{i,x} X + \theta_{i,y} Y + \theta_{i,z} Z$
is a single-qubit observable. The game $G_\theta$ has the payoff
observables
\begin{equation}\label{eq:warmup-games}
\begin{aligned}
  R_1^\theta &= (\theta_1\cdot\vec\sigma)\otimes I\otimes I
     \;-\; \bigl(X\otimes X\otimes I + Y\otimes Y\otimes I + Z\otimes Z\otimes I\bigr),\\
  R_2^\theta &= I\otimes(\theta_2\cdot\vec\sigma)\otimes I
     \;-\; \bigl(I\otimes X\otimes X + I\otimes Y\otimes Y + I\otimes Z\otimes Z\bigr),\\
  R_3^\theta &= I\otimes I\otimes(\theta_3\cdot\vec\sigma)
     \;-\; \bigl(X\otimes I\otimes X + Y\otimes I\otimes Y + Z\otimes I\otimes Z\bigr),
\end{aligned}
\end{equation}
where the three tensor factors are the qubits of players $1$, $2$, and
$3$. Each observable has two terms. The first acts on the player's own
qubit only. The second is the exchange operator $\Gamma$ acting on her own
qubit and on the qubit of her successor, with the identity on the third
qubit; for player $3$, the successor is player $1$. Denote
$(\theta_i\cdot\vec\sigma)_i$ as the operator $\theta_i\cdot\vec\sigma$ on
qubit $i$ and $\Gamma_{i,i+1}$ for $\Gamma$ on qubits $i$ and $i+1$, with
the identity elsewhere in both cases, the three lines
of~\eqref{eq:warmup-games} read
\[
  R_i^\theta = (\theta_i\cdot\vec\sigma)_{i} - \Gamma_{i,\,i+1}
  \qquad (i \in \mathbb{Z}_3).
\]
By~\eqref{eq:warmup-bloch}, applied to each term, the payoffs are
\begin{equation}\label{eq:warmup-payoff}
  u_i^\theta(r_1,r_2,r_3)
  = \theta_i\cdot r_i - r_i\cdot r_{i+1}
  = r_i\cdot(\theta_i - r_{i+1}).
\end{equation}
We write
$\mathcal{G}_\delta = \{\,G_\theta : \|\theta_i\|_2 < \delta
 \text{ for all } i\,\}$
for the family of these games, and we call $G_0$ the \emph{base game}.

\begin{figure}[t]
  \centering
  \includegraphics[width=0.89\linewidth]{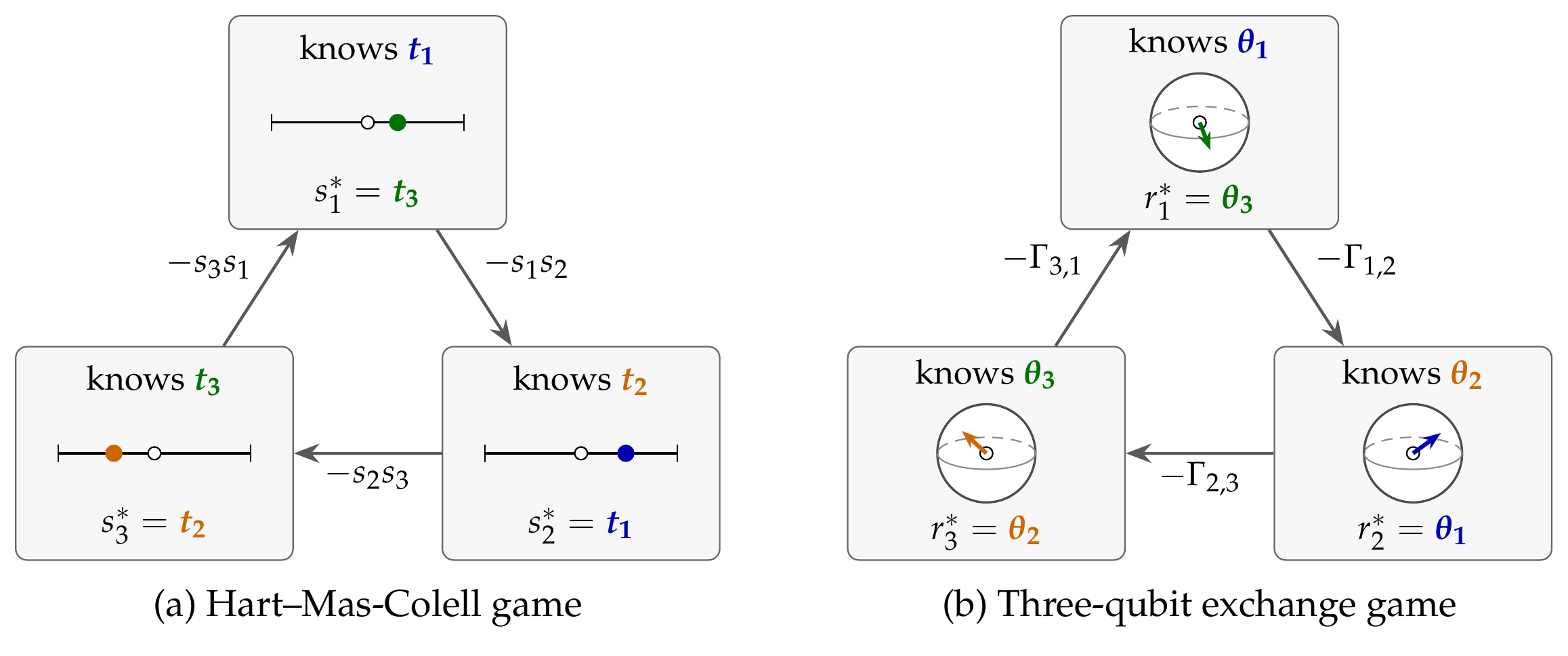}
  \caption{The Hart--Mas-Colell game (a) and the three-qubit exchange game
  (b). Each box is a player, and an arrow to her successor carries the
  payoff term that rewards her for mismatching her successor. A player
  knows only her own parameter, but her equilibrium strategy, which are the colored
  dot on $[-1,1]$ in~(a) and the Bloch vector in~(b), equals the parameter
  of her predecessor and is drawn in the predecessor's color. The white dots mark the equilibrium of the base game, in which all parameters are zero. }
  \label{fig:warmup-hmc}
\end{figure}

The classical family is in fact contained in ours, as the part of the
construction that lies on the $z$-axis. Let $e_z = (0, 0, 1)$. A state
whose Bloch vector lies on the $z$-axis is diagonal,
\[
  P(se_z) = \begin{pmatrix} x & 0 \\ 0 & 1 - x \end{pmatrix}
  \qquad \text{with } s = 2x - 1,
\]
so it is a classical mixed strategy that plays the first action with
probability $x$, and $s$ is its bias. Now restrict the game to the
$z$-axis in two ways. First, choose the parameters on the axis:
$\theta_i = t_i\, e_z$ for real numbers $t_i$. Second, let the players use
diagonal states only: $r_i = s_i\, e_z$ with $s_i \in [-1, 1]$. The
payoff~\eqref{eq:warmup-payoff} then becomes
\[
  u_i^\theta = s_i(t_i - s_{i+1}),
\]
which is the classical payoff of~\eqref{eq:warmup-hmc-bias}. Moreover, the
equilibrium $r_i^* = \theta_{i-1} = t_{i-1}\, e_z$ of
Lemma~\ref{lem:warmup-ne} is itself diagonal, and it is the classical
equilibrium~\eqref{eq:warmup-hmc-ne-bias}. As an illustration, Table \ref{tab:warmup-dictionary} compares the setting between classical and quantum constructions, and Figure~\ref{fig:warmup-hmc} illustrates the cyclic structure of both constructions.

\begin{table}[t]
\centering
\small
\begin{tabular}{@{}>{\raggedright\arraybackslash}p{3.9cm}
                >{\raggedright\arraybackslash}p{5.1cm}
                >{\raggedright\arraybackslash}p{5.6cm}@{}}
\toprule
 & Hart--Mas-Colell~\cite{HartMasColell2003} & This section \\
\midrule
Strategy of player $i$
  & bias $s_i \in [-1, 1]$
  & Bloch vector $r_i \in \mathbb{B}$ \\
\addlinespace
Hidden parameter
  & $t_i \in \R$, with $|t_i|$ small
  & $\theta_i \in \R^3$, with $\|\theta_i\|_2 < \delta$ \\
\addlinespace
Payoff of player $i$
  & $s_i(t_i - s_{i+1})$
  & $r_i\cdot(\theta_i - r_{i+1})$ \\
\addlinespace
Unique equilibrium
  & $s_i^* = t_{i-1}$
  & $r_i^* = \theta_{i-1}$ \\
\addlinespace
Jacobian at the base game
  & $3\times 3$, zero diagonal entries
  & $9\times 9$, zero $3\times 3$ diagonal blocks \\
\bottomrule
\end{tabular}
\caption{The Hart--Mas-Colell construction and its counterpart on the
Bloch ball. The classical column uses the variables $s_i$ and $t_i$ and
the simplified payoff~\eqref{eq:warmup-bias-payoff}.}
\label{tab:warmup-dictionary}
\end{table}

\paragraph{What is different on the Bloch ball.}
Two points require care. First, a classical player has two pure
strategies, whereas a qubit player has a sphere of them, so the uniqueness
of the equilibrium cannot be checked case by case. If the vector
$c_i = \theta_i - r_{i+1}$ is nonzero, the payoff $r_i \cdot c_i$ has the
single maximizer $c_i/\|c_i\|_2$, which is a pure state. Player $i$ can play a mixed
state at equilibrium only if $c_i = 0$, that is, only if her successor
plays exactly $P(\theta_i)$. This is the indifference condition of the
classical game, and it is the reason why the equilibrium of player $i+1$ is
governed by $\theta_i$. To exclude equilibria in pure states we use the
geometry of the ball in place of a finite case analysis. Second, the games
are not classical games in disguise. The operator $\Gamma$ has the
entangled singlet state $\tfrac{1}{\sqrt2}(\ket{01} - \ket{10})$ as its
unique eigenvector of eigenvalue $-3$, so $R_i^0$ is not diagonal in any
product basis. Moreover, $\Gamma$ commutes with $V \otimes V$ for every
unitary $V$, so the base game has no distinguished axis, and the hidden
parameter ranges over a three-dimensional ball of directions and not over
an interval.

\begin{lemma}[Unique Nash equilibrium]\label{lem:warmup-ne}
Let $0 < \delta < \tfrac14$. Every game $G_\theta \in \mathcal{G}_\delta$
has a unique Nash equilibrium. In this equilibrium player $i$ plays
\begin{equation}\label{eq:warmup-ne}
  \rho_i^* = P(\theta_{i-1}),
  \qquad \text{that is,} \qquad
  r_i^* = \theta_{i-1}
  \qquad (i \in \mathbb{Z}_3).
\end{equation}
\end{lemma}

\begin{proof}
Fix $\theta$ and a profile $r$, and let $c_i = \theta_i - r_{i+1}$, so
that $u_i^\theta(r) = r_i\cdot c_i$ by~\eqref{eq:warmup-payoff}. This
payoff is linear in $r_i$. If $c_i \neq 0$, its unique maximizer over
$\mathbb{B}$ is the unit vector $c_i/\|c_i\|_2$. If $c_i = 0$, every
$r_i \in \mathbb{B}$ is a maximizer. The profile $r$ is a Nash equilibrium
if and only if every $r_i$ is such a maximizer.

\noindent\textbf{Existence.} Let $r_i = \theta_{i-1}$ for every $i$. Then
$r_{i+1} = \theta_i$, so $c_i = 0$ for every $i$, and every player is
best-responding.

\noindent\textbf{Uniqueness.} Let $r$ be a Nash equilibrium. Suppose
first that $c_i = 0$ for some $i$. Then $r_{i+1} = \theta_i$, and so
$\|r_{i+1}\|_2 < 1$. If $c_{i+1}$ were nonzero, the unique best response
of player $i+1$ would be a unit vector. Hence $c_{i+1} = 0$. Repeating
the argument around the cycle gives $c_1 = c_2 = c_3 = 0$, that is,
$r_{i+1} = \theta_i$ for every $i$, which is~\eqref{eq:warmup-ne}.

It remains to exclude the case in which $c_i \neq 0$ for every $i$. In
this case every $r_i$ is a unit vector, and $r_i = c_i/\mu_i$ with
$\mu_i = \|\theta_i - r_{i+1}\|_2$. Since $\|r_{i+1}\|_2 = 1$ and
$\|\theta_i\|_2 < \delta$, we have $|\mu_i - 1| < \delta$. Hence
\[
  r_i + r_{i+1}
  = \frac{\theta_i}{\mu_i} + \Bigl(1 - \frac{1}{\mu_i}\Bigr)\, r_{i+1},
  \qquad
  \|r_i + r_{i+1}\|_2
  \le \frac{\|\theta_i\|_2 + |\mu_i - 1|}{\mu_i}
  < \frac{2\delta}{1 - \delta}.
\]
In words, every player is almost exactly opposite to her successor. On a
cycle of odd length this is impossible: since
$2r_1 = (r_1 + r_2) - (r_2 + r_3) + (r_3 + r_1)$,
\[
  2 = \|2r_1\|_2
  \le \|r_1 + r_2\|_2 + \|r_2 + r_3\|_2 + \|r_3 + r_1\|_2
  < \frac{6\delta}{1 - \delta} < 2,
\]
where the last inequality uses $\delta < \tfrac14$. This contradiction
completes the proof.
\end{proof}

Lemma~\ref{lem:warmup-ne} is the qubit counterpart
of~\eqref{eq:warmup-hmc-ne-bias}, and it supplies the second idea of
Section~\ref{sec:warmup-hmc}. Every equilibrium state $P(\theta_{i-1})$ is
mixed, and in the base game every player plays the maximally mixed state
$I/2$. Player $i$ must play exactly the state that makes her predecessor
indifferent, and this state is determined by her predecessor's bias
$\theta_{i-1}$, which does not appear in her own observable $R_i^\theta$.

\subsection{Impossibility for smooth uncoupled dynamics}
\label{sec:warmup-proof}

\paragraph{Dynamics with classical feedback.}
In this section only, we depart from the model of
Definition~\ref{def:prelim-rule} and follow Hart and
Mas-Colell~\cite{HartMasColell2003}. Time is continuous, and in a game
$G = (R_1, R_2, R_3)$ the profile evolves by
\begin{equation}\label{eq:warmup-dynamics}
  \dot r_i = F_i(r_1, r_2, r_3;\, R_i) \qquad (i \in \mathbb{Z}_3),
\end{equation}
which is~\eqref{eq:warmup-hmc-dynamics} in Bloch coordinates. Each update rule $F_i(\,\cdot\,; R_i)$ is continuously differentiable, and
trajectories that start in $\mathbb{B}^3$ remain in $\mathbb{B}^3$, so that
every $r_i(t)$ is the Bloch vector of a state. The dynamics are
\emph{uncoupled}: as the notation indicates, $F_i$ depends on the game
only through player $i$'s own observable $R_i$.

Two features distinguish this model from the one studied in the rest of
the paper. First, the arguments of $F_i$ include the Bloch vectors of the
other players, that is, complete classical descriptions of their states,
and not physical copies of these states. Second, $F_i$ is an arbitrary
smooth function of these vectors and is not required to arise from a
physical operation on the players' qubits. Both features make the players
more powerful, and therefore make the impossibility result stronger.

\begin{definition}[Nash-convergent dynamics~\cite{HartMasColell2003}]
\label{def:warmup-nash-convergent}
Consider a family of games, each with a unique Nash equilibrium. Dynamics
of the form~\eqref{eq:warmup-dynamics} are \emph{Nash-convergent} on the
family if, in every game of the family, the unique Nash equilibrium $r^*$
\begin{enumerate}
  \item is a rest point that attracts every trajectory, and
  \item is \emph{hyperbolically stable}: every eigenvalue of the Jacobian
        of the update rule at $r^*$ has negative real part.
\end{enumerate}
\end{definition}

Condition~2 is the regularity condition of Hart and Mas-Colell. It asks
that the equilibrium attract nearby trajectories at an exponential rate.

\begin{theorem}[Hart--Mas-Colell construction on the Bloch ball]
\label{thm:warmup}
Let $0 < \delta < \tfrac14$. No continuously differentiable uncoupled
dynamics~\eqref{eq:warmup-dynamics} are Nash-convergent on the family
$\mathcal{G}_\delta$. More strongly, if such dynamics have the Nash
equilibrium of every game in $\mathcal{G}_\delta$ as a rest point, then
the Jacobian at the equilibrium of the base game $G_0$ has trace zero, and
hence it has an eigenvalue with nonnegative real part.
\end{theorem}

The proof carries out the third idea of Section~\ref{sec:warmup-hmc}. The
only change is that the hidden parameter now moves the equilibrium of
player $i$ through a three-dimensional ball and not through an interval,
so an entire $3 \times 3$ diagonal block of the Jacobian vanishes.

\begin{proof}
Suppose that uncoupled dynamics~\eqref{eq:warmup-dynamics} have the Nash
equilibrium of every game in $\mathcal{G}_\delta$ as a rest point. Let
$f_i(r) = F_i(r; R_i^0)$ be the update rule of player $i$ in the base
game, and let $f = (f_1, f_2, f_3)$. By Lemma~\ref{lem:warmup-ne}, the
equilibrium of $G_0$ is the origin $r = 0$, which is an interior point of
$\mathbb{B}^3$. Let $J = Df(0)$ be the $9 \times 9$ Jacobian of $f$ at
the origin, with $3 \times 3$ blocks $J_{ij} = D_{r_j} f_i(0)$.

\noindent\textbf{Step 1: the games that player $i$ has the same payoff.} Fix a
player $i$ and a vector $y \in \R^3$ with $\|y\|_2 < \delta$. Let $\theta$
be the parameter with $\theta_{i-1} = y$ and
$\theta_i = \theta_{i+1} = 0$. Since $\theta_i = 0$, player $i$ has the
same observable as in the base game, that is, $R_i^\theta = R_i^0$. By
Lemma~\ref{lem:warmup-ne}, however, the equilibrium of $G_\theta$ is not
the origin. It is the profile $y^{(i)} \in \mathbb{B}^3$ in which player
$i$ plays $y$ and the other two players play $0$.

\noindent\textbf{Step 2: the diagonal blocks vanish.} The profile
$y^{(i)}$ is a rest point of the dynamics in $G_\theta$, so
$F_i(y^{(i)}; R_i^\theta) = 0$. By uncoupledness and Step~1,
\begin{equation}\label{eq:warmup-rest}
  f_i\bigl(y^{(i)}\bigr)
  = F_i\bigl(y^{(i)}; R_i^0\bigr)
  = F_i\bigl(y^{(i)}; R_i^\theta\bigr)
  = 0
  \qquad \text{for every } y \text{ with } \|y\|_2 < \delta .
\end{equation}
Thus $f_i$ vanishes identically when the other two players are at the
origin and player $i$ is anywhere near it. Differentiating
\eqref{eq:warmup-rest} with respect to $y$ at $y = 0$ gives
$J_{ii} = D_{r_i} f_i(0) = 0$. Since $i$ was arbitrary,
\[
  \Tr[J] = \Tr[J_{11}] + \Tr[J_{22}] + \Tr[J_{33}] = 0 .
\]

\noindent\textbf{Step 3: a traceless Jacobian is not stable.} The trace
of $J$ is the sum of its eigenvalues, counted with algebraic
multiplicity. Since the trace is real, the real parts of the eigenvalues
of $J$ sum to zero, so at least one of them is nonnegative. This proves
the second statement. For the first, Nash-convergent dynamics have every
equilibrium as a rest point, and by condition~2 of
Definition~\ref{def:warmup-nash-convergent} every eigenvalue of $J$ has
negative real part. This is a contradiction.
\end{proof}

The proof has a simple interpretation. From the point of view of player
$i$, the profile $y^{(i)}$ may be the equilibrium of a game that she cannot
distinguish from the base game, so she must stay put there, for every
small $y$. Her update rule therefore cannot react to a displacement of
her own state. A system in which no player reacts to her own displacement
has a traceless Jacobian, and it cannot contract in all directions. For
example, near the equilibrium, gradient play
$\dot r_i = \nabla_{r_i} u_i^\theta = \theta_i - r_{i+1}$ is uncoupled and
has every equilibrium of the family as a rest point. Its Jacobian has the
eigenvalues $-1$ and $\tfrac12 \pm \tfrac{\sqrt3}{2}\,\ii$, each with
multiplicity three, so the equilibrium is a saddle from which nearby
trajectories spiral away. Theorem~\ref{thm:warmup} says that no smooth
uncoupled rule can repair this in all games of the family at once.

\begin{remark}[The geometry of the Jacobian]\label{rem:warmup-cyclic}
Here we explore some additional geometric properties of the Jacobian. Let $\theta_i = 0$, $\theta_{i-1} = y$, and
$\theta_{i+1} = w$. Player $i$ still has the observable $R_i^0$, and by
Lemma~\ref{lem:warmup-ne} the equilibrium of $G_\theta$ is the profile in
which player $i$ plays $y$, her predecessor plays $w$, and her successor
plays $0$. Hence $f_i$ vanishes whenever her successor is at the origin
and the other two states are small, so $J_{i,i-1} = 0$ as well as
$J_{ii} = 0$. To first order, player $i$ reacts only to a displacement of
her successor, the one player whose state enters her payoff. The only
nonzero blocks of $J$ are therefore $J_{12}$, $J_{23}$, and $J_{31}$, and
$\det(\lambda I - J) = \det(\lambda^3 I - J_{12}J_{23}J_{31})$. The
eigenvalues of $J$ are thus the cube roots of the eigenvalues of
$J_{12}J_{23}J_{31}$. The three cube roots of a nonzero number are
$120^\circ$ apart, so one of them has positive real part. Consequently,
either $J$ has an eigenvalue with a positive real part or $J$ is nilpotent.
This is the block form of the Jacobian proved by Hart and
Mas-Colell~\cite{HartMasColell2003} in the classical game.
\end{remark}


\section{An Impossibility for Memoryless Quantum Channels}
\label{sec:memoryless-impossibilty}
This section shows that memoryless dynamics cannot converge to Nash
equilibrium, even in a very small family of conjugation games. We consider
discrete-time dynamics of the form~\eqref{eq:prelim-memoryless}, in which
each player updates her strategy by a fixed (autonomous) channel that acts
only on her current strategy and a fresh copy of her opponent's strategy.
We construct four conjugation games in which each player controls a single
qubit and Alice's payoff observable is the same in all four. We then show
that no autonomous, memoryless, uncoupled dynamics converges to Nash
equilibrium from every initial profile in all four games. In fact, any such
dynamics that merely keeps the four equilibria fixed must leave Alice's
strategy unchanged forever. Section~\ref{sec:memoryless-games} presents the
games and explains the mechanism behind the obstruction,
Section~\ref{sec:memoryless-proof} gives the proof, and
Section~\ref{sec:memoryless-nqubit} extends the result to $n$-qubit
registers.
 
\subsection{Four games with conjugate equilibria}\label{sec:memoryless-games}
 
\paragraph{Construction of the games.}
Each player controls one qubit, and each player's strategy space is the
full set $\Dens(\mathbb{C}^2)$. Consider the four pure states $\mathcal{S} = \bigl\{\ket{0},\ \ket{1},\ \ket{+},\ \ket{{+i}}\bigr\},$
\begin{equation}\label{eq:four-states}
  \ket{+} = \tfrac{1}{\sqrt2}(\ket{0} + \ket{1}), \qquad
  \ket{{+i}} = \tfrac{1}{\sqrt2}(\ket{0} + i\ket{1}).
\end{equation}
Let $\ket{\Omega} = \frac{1}{\sqrt{2}}(\ket{00} + \ket{11})$. For each
$\psi \in \mathcal{S}$, the game $G_\psi$ has payoff observables
\begin{equation}\label{eq:four-games}
  R_A = 2\ket{\Omega}\bra{\Omega}, \qquad
  R_{B,\psi} = I \otimes P_\psi .
\end{equation}
By~\eqref{eq:prelim-transpose-payoff}, the payoffs are
\[
  u_A(\rho, \sigma) = \Tr[\rho\,\sigma^\top], \qquad
  u_{B,\psi}(\rho, \sigma) = \Tr[P_\psi\,\sigma].
\]
Bob wants to play the target $P_\psi$, whatever Alice does. Alice wants
to match the transpose of Bob's state. By Lemma~\ref{lem:prelim-conj-ne},
$G_\psi$ has the unique Nash equilibrium
\begin{equation}\label{eq:four-equilibria}
  (\rho_\psi^*, \sigma_\psi^*) = (P_{\overline{\psi}}, P_\psi).
\end{equation}
Explicitly, $\overline{\psi} = \psi$ for $\psi \in \{\ket{0}, \ket{1}, \ket{+}\}$,
and the conjugate of $\ket{{+i}}$ is
$\ket{{-i}} = \frac{1}{\sqrt{2}}(\ket{0} - i\ket{1})$. Thus Alice's equilibrium
strategy coincides with Bob's in three of the games, and is its mirror
image in the fourth.
 
The four games differ only in Bob's payoff. Alice's payoff observable
$R_A$ is the same in all of them, so an uncoupled rule
(Definition~\ref{def:prelim-rule}) gives Alice a \emph{single} update
channel $\Phi_A : \Lin(\mathbb{C}^2 \otimes \mathbb{C}^2) \to \Lin(\mathbb{C}^2)$,
which she must use in every game. The question is whether one physical
channel can steer Alice to $P_{\overline{\psi}}$ in all four games at
once.

\begin{figure}[t]
    \centering
    \includegraphics[width=0.7\linewidth]{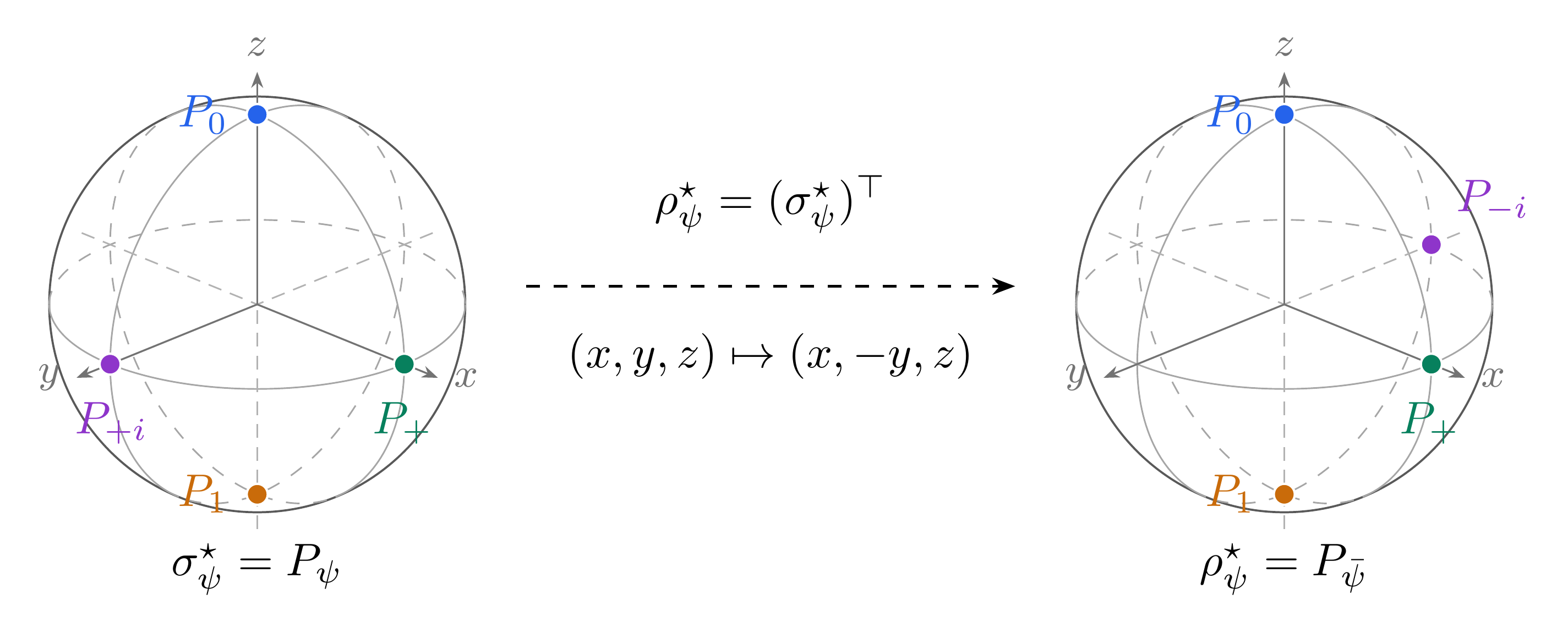}
    \caption{Equilibria of the four-game family.
    For $\psi\in\{0,1,+,+i\}$, write
    $P_\psi=\lvert\psi\rangle\langle\psi\rvert$.
    Bob's unique dominant strategy is $\sigma_\psi^*=P_\psi$,
    and Alice's unique best response is
    $\rho_\psi^*=P_{\bar\psi}=P_\psi^\top$.
    Matching colors identify the same game on both spheres.
    Transposition reflects Bloch vectors across the $xz$-plane,
    fixing $P_0,P_1,P_+$ and sending $P_{+i}$ to $P_{-i}$.
    The dashed arrow depicts the equilibrium relation.}
    \label{fig:four-games-family}
\end{figure}
 
\paragraph{Why these games prevent Nash convergence.}
We now explain informally why no single channel can serve Alice in all four
games. The argument proceeds in three stages: convergence imposes fixed-point
constraints on Alice's channel, these constraints cannot be met by
processing Bob's copy alone, and they can be met only by a channel that
ignores Bob's copy entirely.

The first stage is a direct consequence of continuity. If the
dynamics~\eqref{eq:prelim-memoryless} converge to equilibrium in every
game, then by~\eqref{eq:prelim-fixed-condition} Alice's channel must
satisfy the four fixed-point constraints
\begin{equation}\label{eq:four-constraints}
  \Phi_A\bigl(P_{\overline{\psi}} \otimes P_\psi\bigr) = P_{\overline{\psi}}
  \qquad (\psi \in \mathcal{S}).
\end{equation}
 
For the second stage, suppose that Alice ignores her own register and
computes her next strategy from Bob's copy alone, so that
$\Phi_A(\rho \otimes \sigma) = \Lambda(\sigma)$ for some channel $\Lambda$. Then~\eqref{eq:four-constraints} would require
$\Lambda(P_\psi) = P_\psi^\top$ for all four $\psi \in \mathcal{S}$. The identities
\[
  P_0 + P_1 = I, \qquad P_0 - P_1 = Z, \qquad
  2P_+ - I = X, \qquad 2P_{+i} - I = Y
\]
show that the four projectors span $\Lin(\mathbb{C}^2)$. Hence
$\Lambda$ would agree with the transpose map on every matrix, which is
impossible because the transpose is not completely positive. In other
words, Alice cannot learn her answer from Bob's copy.

The third stage shows that combining both inputs does not help either.
Alice also holds her own current strategy, and at equilibrium that
strategy already equals the answer $P_{\overline{\psi}}$. The channel
$\Phi_A = \Tr_B$, which keeps Alice's register and discards Bob's,
therefore satisfies~\eqref{eq:four-constraints}
(Remark~\ref{rem:prelim-not-trivial}). Lemma \ref{lem:four-state} below shows that
it is the \emph{only} channel that does this. The reason is as follows. Each
constraint asks for a \emph{pure} output, so by
Lemma~\ref{lem:prelim-pure-output} everything that the channel extracts
from Bob's copy $\ket{\psi}$ must be deposited in the environment, as a
vector $\ket{e_\psi}$. Inner-product preservation then forces these
environment vectors to have the same overlaps as Bob's states. In other
words, the environment holds an isometric copy $W\ket{\psi}$ of Bob's
input, and no information about $\psi$ reaches Alice's register.
Because the four constraint inputs
$\ket{\overline{\psi}} \otimes \ket{\psi}$ span the entire two-qubit input
space, these conditions determine the channel completely, and it is
$\Tr_B$. Alice's strategy then never changes, so the dynamics cannot
converge from any initial state other than the equilibrium itself.
 
\paragraph{Essential properties of the games.}
Three features of the family are used in the argument. First, the targets
are \emph{pure}, which allows us to factor the Stinespring dilation and
isolate an environment vector for each game. Second, most pairs of targets
are \emph{nonorthogonal}, which allows us to cancel the common factor
$\overline{\braket{\psi}{\phi}}$ in the overlap identity and thereby
transfer Bob's overlaps to the environment. The only orthogonal pair,
$\ket{0}$ and $\ket{1}$, is handled through the nonzero overlaps of both
states with $\ket{+}$ and $\ket{{+i}}$. Third, the \emph{complex} state
$\ket{{+i}}$ makes the constraints span the whole input space. If $\psi$ is
real, then $\ket{\overline{\psi}} \otimes \ket{\psi} = \ket{\psi} \otimes
\ket{\psi}$ lies in the three-dimensional symmetric subspace. The input
\[
  \ket{{-i}} \otimes \ket{{+i}}
  = \tfrac12\bigl(\ket{00} + \ket{11} + i(\ket{01} - \ket{10})\bigr)
\]
is the only one with a component along the antisymmetric vector
$\ket{01} - \ket{10}$, and without it the channel would be unconstrained in
that direction. Indeed, for real targets the rule in which Alice copies
Bob's register converges within two rounds
(Remark~\ref{rem:prelim-not-trivial}).

The obstruction is also genuinely quantum. The observable $R_A$ has rank
one with an entangled eigenvector, so it is not diagonal in any product
basis and does not reduce to a classical bimatrix game.
 
\subsection{Impossibility theorem for Nash convergence}
\label{sec:memoryless-proof}
 
We first show that the four constraints~\eqref{eq:four-constraints}
determine Alice's channel, and then prove the impossibility theorem.
 
\begin{lemma}[Four-state rigidity]\label{lem:four-state}
Let $\Phi : \Lin(\mathbb{C}^2 \otimes \mathbb{C}^2) \to \Lin(\mathbb{C}^2)$
be a quantum channel such that
\begin{equation}\label{eq:rigidity-hyp}
  \Phi\bigl(P_{\overline{\psi}} \otimes P_\psi\bigr) = P_{\overline{\psi}}
  \qquad \text{for every } \psi \in \mathcal{S}.
\end{equation}
Then $\Phi = \Tr_B$. In particular, $\Phi(\rho \otimes \sigma) = \rho$ for
all density matrices $\rho$ and $\sigma$.
\end{lemma}
 
\begin{proof}
By Theorem~\ref{thm:prelim-stinespring}, there are a finite-dimensional
environment $E$ and an isometry
$U : \mathbb{C}^2 \otimes \mathbb{C}^2 \to \mathbb{C}^2 \otimes E$ with
$\Phi(X) = \Tr_E[U X U^\dagger]$. We determine $U$ in the following steps.
 
\noindent\textbf{Step 1: environment vectors.} For each $\psi \in \mathcal{S}$, the
input $\ket{\overline{\psi}} \otimes \ket{\psi}$ is a pure state, and by
\eqref{eq:rigidity-hyp} its output $P_{\overline{\psi}}$ is pure.
Lemma~\ref{lem:prelim-pure-output} gives a unit vector $e_\psi \in E$ with
\begin{equation}\label{eq:env-vectors}
  U\bigl(\ket{\overline{\psi}} \otimes \ket{\psi}\bigr)
  = \ket{\overline{\psi}} \otimes \ket{e_\psi}.
\end{equation}
The vector $e_\psi$ is the state left in the environment when the
channel acts in game $G_\psi$.
 
\noindent\textbf{Step 2: environment overlaps equal Bob's overlaps.} Let
$\psi, \phi \in \mathcal{S}$. Since
$\braket{\overline{\psi}}{\overline{\phi}} = \overline{\braket{\psi}{\phi}}$,
the input overlap is
$\overline{\braket{\psi}{\phi}}\,\braket{\psi}{\phi}$ and the output
overlap is $\overline{\braket{\psi}{\phi}}$. Equation~\eqref{eq:prelim-overlap}
therefore becomes
\[
  \overline{\braket{\psi}{\phi}}\,\braket{\psi}{\phi}
  = \overline{\braket{\psi}{\phi}}\,\braket{e_\psi}{e_\phi}.
\]
Whenever $\braket{\psi}{\phi} \neq 0$, we may cancel this factor from both
sides and obtain
\begin{equation}\label{eq:env-overlaps}
  \braket{e_\psi}{e_\phi} = \braket{\psi}{\phi}.
\end{equation}
In $\mathcal{S}$, the only orthogonal pair is $\{\ket{0}, \ket{1}\}$. Thus every
other pair has a nonzero overlap:
\[
  \braket{0}{+} = \braket{0}{{+i}} = \braket{1}{+} = \tfrac{1}{\sqrt2},
  \qquad
  \braket{1}{{+i}} = \tfrac{i}{\sqrt2},
  \qquad
  \braket{+}{{+i}} = \tfrac{1+i}{2}.
\]
Hence~\eqref{eq:env-overlaps} holds for every pair except $\{0, 1\}$.
 
\noindent\textbf{Step 3: produce the environment vectors with an isometry.} The
three states $\ket{0}, \ket{+}, \ket{{+i}}$ are pairwise nonorthogonal, so
by~\eqref{eq:env-overlaps} the vectors $e_0, e_+, e_{+i}$ have the same
Gram matrix as $\ket{0}, \ket{+}, \ket{{+i}}$. These three states span
$\mathbb{C}^2$. By Lemma~\ref{lem:prelim-gram}, there is a linear isometry
$W : \mathbb{C}^2 \to E$ with
\[
  W\ket{\psi} = \ket{e_\psi} \qquad \text{for } \psi \in \{0, +, {+i}\}.
\]
It remains to show that $W\ket{1} = \ket{e_1}$, even though $\ket{1}$ is
orthogonal to $\ket{0}$. For $s \in \{+, {+i}\}$,
using~\eqref{eq:env-overlaps} for the nonorthogonal pair $(s, 1)$, we get
\[
  \bra{s} W^\dagger \ket{e_1} = \braket{e_s}{e_1} = \braket{s}{1}.
\]
Since $\ket{+}$ and $\ket{{+i}}$ span $\mathbb{C}^2$, this gives
$W^\dagger\ket{e_1} = \ket{1}$ and
$\bra{1}W^\dagger\ket{e_1} = 1$. Both $W\ket{1}$ and $\ket{e_1}$ are unit
vectors, so
\[
  \bigl\| \ket{e_1} - W\ket{1} \bigr\|^2
  = 2 - 2\,\mathrm{Re}\,\bra{1}W^\dagger\ket{e_1} = 0 .
\]
Therefore $\ket{e_\psi} = W\ket{\psi}$ for every $\psi \in \mathcal{S}$.
 
\noindent \textbf{Step 4: determine the dilation.} By~\eqref{eq:env-vectors} and
Step~3, for every $\psi \in \mathcal{S}$, we have
\[
  U\bigl(\ket{\overline{\psi}} \otimes \ket{\psi}\bigr)
  = \ket{\overline{\psi}} \otimes W\ket{\psi}
  = (I \otimes W)\bigl(\ket{\overline{\psi}} \otimes \ket{\psi}\bigr).
\]
The four input vectors are
\[
  \ket{00}, \qquad \ket{11}, \qquad
  \tfrac12\bigl(\ket{00} + \ket{01} + \ket{10} + \ket{11}\bigr), \qquad
  \tfrac12\bigl(\ket{00} + i\ket{01} - i\ket{10} + \ket{11}\bigr).
\]
After removing their $\ket{00}$ and $\ket{11}$ components, the last two vectors
become \[\tfrac12(\ket{01} + \ket{10}),\qquad\tfrac{i}{2}(\ket{01} - \ket{10}),\]
which are linearly independent. So the four vectors form a basis of
$\mathbb{C}^2 \otimes \mathbb{C}^2$. Two linear maps that agree on a basis
are equal, so $U = I \otimes W$. Finally, for all $Y, Z \in \Lin(\mathbb{C}^2)$, we get
\[
  \Phi(Y \otimes Z)
  = \Tr_E\bigl[(I \otimes W)(Y \otimes Z)(I \otimes W^\dagger)\bigr]
  = Y\,\Tr[W Z W^\dagger]
  = \Tr[Z]\, Y ,
\]
using the cyclicity of the trace and $W^\dagger W = I$. By linearity,
$\Phi = \Tr_B$.
\end{proof}
 
The proof shows that the channel deposits an exact copy $W\ket{\psi}$ of
Bob's input in the environment. The copy is not conjugated, and none of
the information it carries reaches Alice's register.

 \begin{figure}[t]
    \centering
    \includegraphics[width=0.8\linewidth]{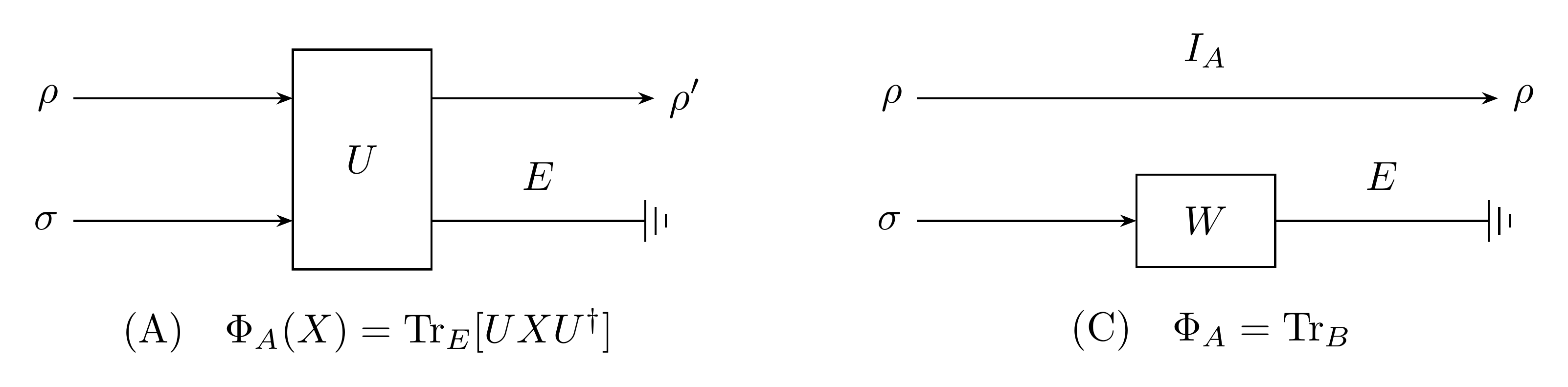}
    \caption{A general Stinespring dilation (A) and the structure
    forced by preserving all four equilibria (C).
    In (A), $U$ is an isometry and
    $\rho'=\Phi_A(\rho\otimes\sigma)$.
    The four conditions
    $\Phi_A(P_{\bar\psi}\otimes P_\psi)=P_{\bar\psi}$ for
    $\psi\in\{0,1,+,+i\}$ imply
    $U=I_A\otimes W$ for an isometry $W$.
    After discarding $E$, the resulting channel is
    $\Phi_A=\operatorname{Tr}_B$, which traces out Bob's system.
    Thus $\rho_{t+1}=\rho_t$ regardless of Bob's state.}
    \label{fig:dilation-rigidity}
\end{figure}

\begin{theorem}[Impossibility for memoryless dynamics]\label{thm:four-game}
No autonomous, memoryless, uncoupled dynamics~\eqref{eq:prelim-memoryless}
converge to Nash equilibrium from every initial strategy profile in all
four games $\{G_\psi : \psi \in \mathcal{S}\}$. More strongly, if such
dynamics leave the Nash equilibrium of every $G_\psi$ fixed, then Alice's
channel is $\Tr_B$, and $\rho_t = \rho_0$ along every trajectory of every
game in the family.
\end{theorem}
 
\begin{proof}
Suppose that Alice and Bob use autonomous memoryless rules, assigned
uncoupledly to the four games. Since $R_A$ is the same in every game,
Alice uses the same channel $\Phi_A$ in all four. Bob's channels
$\Psi_B$ may differ from game to game and are otherwise arbitrary.
 
First, assume that the equilibrium of every $G_\psi$ is a fixed point of
the dynamics. Then $\Phi_A$ satisfies~\eqref{eq:four-constraints}, which
is the hypothesis of Lemma~\ref{lem:four-state}. Hence $\Phi_A = \Tr_B$,
and for every game and every $t$,
\[
  \rho_{t+1} = \Tr_B[\rho_t \otimes \sigma_t] = \rho_t .
\]
Thus Alice's strategy remains $\rho_0$ forever, regardless of Bob's
behavior. This proves the second statement.
 
Now assume, for contradiction, that the dynamics converge to Nash
equilibrium from every initial profile in all four games. Start a
trajectory of $G_\psi$ at the profile $(P_{\overline{\psi}}, P_\psi)$. It
converges to the unique equilibrium of $G_\psi$, and since channels are
continuous, the limit of a trajectory of~\eqref{eq:prelim-memoryless} is
a fixed point. So every equilibrium is fixed, and by the first part
Alice's strategy never changes. Choose any game $G_\psi$ and any initial
profile with $\rho_0 \neq P_{\overline{\psi}}$. Along this trajectory
$\rho_t = \rho_0 \neq P_{\overline{\psi}}$ for all $t$, so the trajectory
does not converge to the unique equilibrium~\eqref{eq:four-equilibria}.
This contradiction proves the theorem.
\end{proof}
 
\begin{remark}[Scope of the result]
The theorem places no restriction on Bob's rules, and it uses only two
properties of Alice's: she applies one physical channel throughout the
family, and that channel acts on her strategy register and a fresh copy
of Bob's strategy, with no further persistent memory. The conclusion is
not specific to asymptotic convergence. Any memoryless rule whose fixed
points include all four equilibria leaves Alice's strategy rigid.
\end{remark}

\begin{remark}[Relation to other works]
The factorization and inner-product steps in this proof follow the argument
behind the no-cloning theorem~\cite{WoottersZurek1982,Dieks1982} and the
no-programming theorem of Nielsen and Chuang~\cite{NielsenChuang1997}; see
\cite{KoashiImoto2002,KretschmannSchlingemannWerner2008} for general structure
results on operations that leave a set of states undisturbed.    
\end{remark}

\subsection{Extension to $n$-qubit registers}\label{sec:memoryless-nqubit}
 
The four-game family uses one qubit per player. We now show that the same
obstruction holds when each player controls $n$ qubits, for every
$n \ge 1$, and even when the players' channels act collectively on all of
their qubits. The construction takes tensor products of the four
single-qubit targets, which produces $4^n$ games, and the proof of
Lemma~\ref{lem:four-state} carries over almost verbatim. Only two steps
require new care: finding a large family of pairwise nonorthogonal targets,
and handling targets that are orthogonal to some members of that family.
 
Throughout this subsection, $d = 2^n$, and each player's register is
$\mathbb{C}^d = (\mathbb{C}^2)^{\otimes n}$. The lower case
$\psi, \tau$ denote single-qubit-states, and uppercase
$\Psi, \Sigma$ denote $n$-qubit-states.
 
\paragraph{The construction of games.}
Let $\mathcal{S}$ be the four single-qubit-states
in~\eqref{eq:four-states}. The $n$-qubit targets are all tensor products
of these states:
\[
  \mathcal{S}_n
  = \bigl\{\, \Psi = \psi_1 \otimes \psi_2 \otimes \cdots \otimes \psi_n
      \;:\; \psi_i \in \mathcal{S} \text{ for each } i \,\bigr\}.
\]
There are $4^n = d^2$ targets. Conjugation acts factor by factor, so
$\overline{\Psi} = \overline{\psi_1} \otimes \cdots \otimes \overline{\psi_n}$.
Let $\ket{\Omega_d} = \tfrac{1}{\sqrt{d}} \sum_{a=0}^{d-1} \ket{a} \otimes \ket{a}$.
For each $\Psi \in \mathcal{S}_n$, the game $G_\Psi$ has payoff observables
\begin{equation}\label{eq:nqubit-games}
  R_A = d\, \ket{\Omega_d}\bra{\Omega_d}, \qquad
  R_{B,\Psi} = I \otimes P_\Psi .
\end{equation}
As in the one-qubit case,
$u_A(\rho, \sigma) = \Tr[\rho\, \sigma^\top]$ and
$u_{B,\Psi}(\rho, \sigma) = \Tr[P_\Psi\, \sigma]$, and
Lemma~\ref{lem:prelim-conj-ne} shows that $G_\Psi$ has the unique Nash
equilibrium $(P_{\overline{\Psi}}, P_\Psi)$. For $n = 1$, this is exactly
the family of Section~\ref{sec:memoryless-games}.
 
Only the \emph{targets} have product form. Each player's strategy space
is the full set $\Dens(\mathbb{C}^d)$, which includes states that are
entangled across the player's own $n$ qubits, and each update channel may
act collectively on all of its input qubits. Since $R_A$ is the same in
all $4^n$ games, an uncoupled rule again gives Alice a single channel
$\Phi_A : \Lin(\mathbb{C}^d \otimes \mathbb{C}^d) \to \Lin(\mathbb{C}^d)$.
 
\paragraph{What changes in the proof.}
The proof of Lemma~\ref{lem:four-state} has three ingredients: an overlap
identity for nonorthogonal targets, a single isometry $W$ built from a
pairwise nonorthogonal family that spans Bob's register, and a spanning
argument for the constraint inputs. The overlap identity holds verbatim,
since it uses only the purity of the outputs and the fact that the dilation
is an isometry. For the isometry, the one-qubit proof used the pairwise
nonorthogonal family $\{\ket{0}, \ket{+}, \ket{{+i}}\}$; for $n$ qubits we
use all tensor products of these three states, whose overlaps are products
of nonzero single-qubit overlaps and hence nonzero. A new difficulty arises
for targets that contain $\ket{1}$ factors, because such targets are
orthogonal to some of these reference vectors. For each such target we
choose a basis of reference vectors that avoids the problem, by using
$\ket{+}$ and $\ket{{+i}}$ in the positions where the target has a
$\ket{1}$. Finally, the spanning argument reduces to the one-qubit case:
after the qubits are reordered so that Alice's $i$-th qubit sits next to
Bob's $i$-th qubit, each constraint input is a tensor product of the four
two-qubit vectors from the proof of Lemma~\ref{lem:four-state}, and tensor
products of bases are bases.
 
\begin{lemma}[Rigidity for $n$-qubit registers]\label{lem:nqubit-rigidity}
Let $\Phi : \Lin(\mathbb{C}^d \otimes \mathbb{C}^d) \to \Lin(\mathbb{C}^d)$
be a quantum channel such that
\begin{equation}\label{eq:nqubit-hyp}
  \Phi\bigl(P_{\overline{\Psi}} \otimes P_\Psi\bigr) = P_{\overline{\Psi}}
  \qquad \text{for every } \Psi \in \mathcal{S}_n.
\end{equation}
Then $\Phi = \Tr_B$.
\end{lemma}
 
\begin{proof}
By Theorem~\ref{thm:prelim-stinespring}, write
$\Phi(X) = \Tr_E[U X U^\dagger]$ for an isometry
$U : \mathbb{C}^d \otimes \mathbb{C}^d \to \mathbb{C}^d \otimes E$. We
follow the four steps of the proof of Lemma~\ref{lem:four-state}.
 
\noindent\textbf{Step 1: environment vectors.} Each output
in~\eqref{eq:nqubit-hyp} is pure, so Lemma~\ref{lem:prelim-pure-output}
gives a unit vector $e_\Psi \in E$ with
\begin{equation}\label{eq:nqubit-env}
  U\bigl(\ket{\overline{\Psi}} \otimes \ket{\Psi}\bigr)
  = \ket{\overline{\Psi}} \otimes \ket{e_\Psi}
  \qquad (\Psi \in \mathcal{S}_n).
\end{equation}
 
\noindent\textbf{Step 2: the overlap identity.} For $\Psi, \Sigma \in \mathcal{S}_n$,
we have $\braket{\overline{\Psi}}{\overline{\Sigma}}
= \overline{\braket{\Psi}{\Sigma}}$. Equation~\eqref{eq:prelim-overlap}
therefore reads
$\overline{\braket{\Psi}{\Sigma}}\,\braket{\Psi}{\Sigma}
 = \overline{\braket{\Psi}{\Sigma}}\,\braket{e_\Psi}{e_\Sigma}$, and
cancellation gives
\begin{equation}\label{eq:nqubit-overlaps}
  \braket{e_\Psi}{e_\Sigma} = \braket{\Psi}{\Sigma}
  \qquad \text{whenever } \braket{\Psi}{\Sigma} \neq 0 .
\end{equation}
 
\noindent\textbf{Step 3: a single isometry on a reference family.} Consider the
reference family
\[
  \mathcal{F}_n
  = \bigl\{\, \tau_1 \otimes \cdots \otimes \tau_n
      \;:\; \tau_r \in \{\ket{0}, \ket{+}, \ket{{+i}}\} \,\bigr\}
  \subset \mathcal{S}_n .
\]
For $\Sigma = \tau_1 \otimes \cdots \otimes \tau_n$ and
$\Sigma' = \tau'_1 \otimes \cdots \otimes \tau'_n$ in $\mathcal{F}_n$,
\[
  \braket{\Sigma}{\Sigma'} = \prod_{r=1}^n \braket{\tau_r}{\tau'_r} \neq 0,
\]
because any two of $\ket{0}, \ket{+}, \ket{{+i}}$ have nonzero overlap
(Step~2 of the proof of Lemma~\ref{lem:four-state}). Hence, by
\eqref{eq:nqubit-overlaps}, the vectors $\{e_\Sigma : \Sigma \in \mathcal{F}_n\}$
have the same Gram matrix as $\mathcal{F}_n$. Moreover, $\mathcal{F}_n$
spans $\mathbb{C}^d$, since it contains every tensor product of the
basis $\{\ket{0}, \ket{+}\}$ of $\mathbb{C}^2$. By
Lemma~\ref{lem:prelim-gram}, there is a linear isometry
$W : \mathbb{C}^d \to E$ with
\begin{equation}\label{eq:nqubit-W}
  W\ket{\Sigma} = \ket{e_\Sigma}
  \qquad \text{for every } \Sigma \in \mathcal{F}_n .
\end{equation}
 
\noindent\textbf{Step 4a: every target is handled by the same isometry.} Fix
$\Psi = \psi_1 \otimes \cdots \otimes \psi_n \in \mathcal{S}_n$. We show
that $e_\Psi = W\Psi$. For each position $i$, choose a two-element basis
of $\mathbb{C}^2$ whose members both have nonzero overlap with $\psi_i$:
\[
  \mathcal{B}_i =
  \begin{cases}
    \{\ket{+}, \ket{{+i}}\} & \text{if } \psi_i = \ket{1}, \\[2pt]
    \{\ket{0}, \ket{+}\}    & \text{otherwise.}
  \end{cases}
\]
This choice works because $\ket{1}$ is orthogonal only to $\ket{0}$, while
each of $\ket{0}, \ket{+}, \ket{{+i}}$ has nonzero overlap with both
$\ket{0}$ and $\ket{+}$. Let $\mathcal{B}_\Psi$ be the set of $2^n$
tensor products with $i$-th factor in $\mathcal{B}_i$. It is a basis of
$\mathbb{C}^d$, it is contained in $\mathcal{F}_n$, and each of its
members $\Sigma$ satisfies $\braket{\Sigma}{\Psi} \neq 0$. For
$\Sigma \in \mathcal{B}_\Psi$, \eqref{eq:nqubit-W}
and~\eqref{eq:nqubit-overlaps} give
\[
  \bra{\Sigma} W^\dagger \ket{e_\Psi}
  = \braket{e_\Sigma}{e_\Psi}
  = \braket{\Sigma}{\Psi}.
\]
Since $\mathcal{B}_\Psi$ is a basis, $W^\dagger\ket{e_\Psi} = \ket{\Psi}$,
and so $\bra{\Psi} W^\dagger \ket{e_\Psi} = 1$. Both $W\ket{\Psi}$ and
$\ket{e_\Psi}$ are unit vectors, so
\[
  \bigl\| \ket{e_\Psi} - W\ket{\Psi} \bigr\|^2
  = 2 - 2\,\mathrm{Re}\,\bra{\Psi} W^\dagger \ket{e_\Psi} = 0 .
\]
Therefore $\ket{e_\Psi} = W\ket{\Psi}$ for every $\Psi \in \mathcal{S}_n$.
 
\noindent\textbf{Step 4b: determine the dilation.} By~\eqref{eq:nqubit-env}
and Step~4a,
$U(\ket{\overline{\Psi}} \otimes \ket{\Psi})
 = (I \otimes W)(\ket{\overline{\Psi}} \otimes \ket{\Psi})$ for every
$\Psi \in \mathcal{S}_n$. It remains to check that these $4^n$ input
vectors form a basis of $\mathbb{C}^d \otimes \mathbb{C}^d$. Reorder the
$2n$ qubits so that Alice's $i$-th qubit is followed by Bob's $i$-th
qubit. This reordering is a fixed unitary, so it preserves linear
independence, and it sends
\[
  \ket{\overline{\Psi}} \otimes \ket{\Psi}
  \;\longmapsto\;
  \bigl(\ket{\overline{\psi_1}} \otimes \ket{\psi_1}\bigr)
  \otimes \cdots \otimes
  \bigl(\ket{\overline{\psi_n}} \otimes \ket{\psi_n}\bigr).
\]
As $\Psi$ ranges over $\mathcal{S}_n$, each factor
$\ket{\overline{\psi_i}} \otimes \ket{\psi_i}$ ranges independently over
the four two-qubit vectors that form a basis of
$\mathbb{C}^2 \otimes \mathbb{C}^2$ in the proof of
Lemma~\ref{lem:four-state}. Tensor products of bases form a basis, so
the $4^n$ input vectors are a basis. Hence $U = I \otimes W$, and, exactly
as in the one-qubit case,
$\Phi(Y \otimes Z) = \Tr[W Z W^\dagger]\, Y = \Tr[Z]\, Y$. By linearity,
$\Phi = \Tr_B$.
\end{proof}
 
For $n = 1$, the reference family is $\{\ket{0}, \ket{+}, \ket{{+i}}\}$,
the basis chosen for the target $\ket{1}$ is $\{\ket{+}, \ket{{+i}}\}$, and
the proof reduces to that of Lemma~\ref{lem:four-state}.
 
\begin{theorem}[Impossibility for $n$-qubit registers]\label{thm:nqubit}
For every $n \ge 1$, no autonomous, memoryless, uncoupled
dynamics~\eqref{eq:prelim-memoryless} converge to Nash equilibrium from
every initial strategy profile in all $4^n$ games
$\{G_\Psi : \Psi \in \mathcal{S}_n\}$ defined by~\eqref{eq:nqubit-games}.
This holds even though each player's channel may act collectively on all
$2n$ input qubits. More strongly, if such dynamics leave every
equilibrium of the family fixed, then Alice's channel is $\Tr_B$, and her
strategy never changes along any trajectory.
\end{theorem}
 
\begin{proof}
The argument is that of Theorem~\ref{thm:four-game}, with
Lemma~\ref{lem:nqubit-rigidity} in place of Lemma~\ref{lem:four-state}.
Since $R_A$ is common to all $4^n$ games, uncoupledness gives Alice a
single channel $\Phi_A$. If every equilibrium
$(P_{\overline{\Psi}}, P_\Psi)$ is a fixed point, then $\Phi_A$
satisfies~\eqref{eq:nqubit-hyp}, so $\Phi_A = \Tr_B$ and
$\rho_{t+1} = \rho_t$ for all $t$. If the dynamics converged to
equilibrium from every initial profile, then, by continuity, every
equilibrium would be a fixed point. Alice's strategy would therefore be
frozen, and any trajectory of $G_\Psi$ started with
$\rho_0 \neq P_{\overline{\Psi}}$ would fail to converge, a
contradiction.
\end{proof}
 
\begin{remark}[Size of the family and strength of the conclusion]
The family has $4^n = d^2$ games, which is the smallest number for which
the present argument can work: each game contributes one constraint input
$\ket{\overline{\Psi}} \otimes \ket{\Psi}$, and these inputs must span the
$d^2$-dimensional space $\mathbb{C}^d \otimes \mathbb{C}^d$ for the
constraints to determine $U$. Although the targets are product states, the
conclusion is not restricted to product strategies or local channels. It
applies to arbitrary initial density matrices on $\mathbb{C}^d$, including
states entangled across a player's own qubits, and to channels that process
all qubits jointly.
\end{remark}

\section{Finite Memory Overcomes the Finite-Games Barrier}
\label{sec:finite-classical}
 
Section~\ref{sec:memoryless-impossibilty} showed that a memoryless channel
cannot steer Alice to equilibrium in all four games, nor in their
$4^n$-game extension. In this section we allow Alice a small
\emph{classical} memory and show that the obstruction disappears: a
logarithmic number of memory bits suffices to reach equilibrium in every
game of any finite family of conjugation games.
Section~\ref{sec:finite-classical-overview} describes Alice's strategy
informally and explains why it escapes the rigidity of
Lemma~\ref{lem:four-state}. Section~\ref{sec:finite-classical-formal}
expresses the strategy as a precise update channel and reduces the
resulting dynamics to a Markov chain on labels.
Section~\ref{sec:finite-classical-theorem} bounds the mixing time of this
chain and derives convergence to Nash equilibrium.

\subsection{How finite memory circumvents the obstruction}
\label{sec:finite-classical-overview}

Without memory, Alice faces an impossible task: she receives a physical
copy of Bob's unknown state $\ket{\psi}$ and must turn it into
$\ket{\overline{\psi}}$, which amounts to transposing an unknown state.
The key observation is that Alice does not need to transform Bob's state
at all. She knows the finite list of possible targets in advance. If she
can work out which target Bob is playing, she can prepare the
corresponding answer from its classical description, which is an
ordinary preparation. Memory lets her carry out this identification over
many rounds. The task therefore changes from transforming an unknown
state, which is impossible, to \emph{identifying} one of finitely many
known states, which is possible given enough copies.
 
\paragraph{Alice's strategy.}
Alice keeps in her memory a classical label $k$, which represents her
current guess of Bob's target. In each round she first \emph{verifies} the
guess: she asks the yes/no question ``is Bob's state the target $P_k$?''\
by measuring the received copy with the two-outcome measurement
$\{P_k,\, I - P_k\}$. She then \emph{updates} the guess, keeping $k$ if the
answer is yes and moving on to the next label in a fixed cyclic order if
the answer is no. Finally, she \emph{plays} the best response to her
current guess: she discards her old strategy and prepares $P_k^\top$ for
the updated label $k$. Bob, for his part, plays his dominant target $P_j$
in every round.

\paragraph{Why the strategy reaches equilibrium.}
Convergence rests on two facts about the test. The correct guess is never
rejected: if $k = j$, then Bob's state is exactly $P_j$, so the test
answers yes with probability one, and once Alice reaches the correct label
she keeps it forever and plays her equilibrium strategy $P_j^\top$.
Conversely, every wrong guess is eventually rejected: if $k \neq j$, the
test answers no with probability $1 - \Tr[P_k P_j] > 0$, because the
targets are distinct pure states. Since Alice tests a fresh copy in each
round, the tests behave like independent biased coin flips, and a wrong
guess is rejected after finitely many rounds with probability one. As
Alice cycles through the labels, she therefore reaches the correct label
after finitely many rounds with probability one. For example, in the
four-game family, if Bob plays $\ket{{+i}}$ and Alice guesses $\ket{0}$,
each test rejects her guess with probability
$1 - |\braket{0}{{+i}}|^2 = \tfrac12$.
 
\paragraph{Why this is uncoupled.}
Alice's rule depends only on the fixed list of targets, which is the
same for the whole family, and never on which game is being played. Bob's
rule depends only on his own payoff. Both players use the same channel
in every round, so the rules are also autonomous.
 
\paragraph{Why the rigidity lemma no longer applies.}
The proof of Lemma~\ref{lem:four-state} needed the equilibrium inputs of
different games to be \emph{nonorthogonal}. Their overlaps were cancelled
to show that Bob's information must end up in the environment. With
memory, Alice's persistent state at the equilibrium of game $j$ is
$\ket{j}\bra{j} \otimes P_j^\top$: the label sits next to her strategy.
Labels of different games are orthogonal, so the equilibrium states of
different games are orthogonal \emph{even when their strategy parts
overlap}. For orthogonal inputs, the overlap identity becomes $0 = 0$ and
gives no information. Moreover, these inputs no longer span Alice's
input space. The equilibrium constraints therefore no longer pin down
the channel, and the guess-and-test channel becomes possible. Put
simply, the classical label turns a quantum distinction (nonorthogonal
states) into a classical one (orthogonal labels), and Bob's fresh copy in
each round supplies the evidence that lets the label settle on the
truth.
 
\subsection{Alice's strategy as a Markov chain}\label{sec:finite-classical-formal}
In this subsection we express Alice's strategy as an explicit quantum
channel and show that, after the first round, the resulting dynamics are
fully described by a Markov chain on her labels. The convergence analysis
of Section~\ref{sec:finite-classical-theorem} then reduces to bounding the
mixing time of this chain.

\paragraph{The game family.}
Fix $d \ge 2$ and $L \ge 2$ distinct rank-one projectors
$P_1, \dots, P_L$ on $\mathbb{C}^d$ (the \emph{targets}). For each
$j \in \{1, \dots, L\}$, the game $G_j$ has payoff observables
\begin{equation}\label{eq:fc-games}
  R_A = d\, \ket{\Omega_d}\bra{\Omega_d}, \qquad
  R_B^j = I \otimes P_j .
\end{equation}
By Lemma~\ref{lem:prelim-conj-ne}, $G_j$ has the unique Nash equilibrium
$(P_j^\top, P_j)$, where $P_j^\top$ is again a rank-one projector. The
four-game family of Section~\ref{sec:memoryless-games} is the case $d = 2$,
$L = 4$, and the $n$-qubit family of
Section~\ref{sec:memoryless-nqubit} is the case $d = 2^n$, $L = 4^n$.
 
For labels $k$ and $j$, let
\begin{equation}\label{eq:fc-q}
  q_{kj} = 1 - \Tr[P_k P_j]
\end{equation}
be the probability that the test $\{P_k, I - P_k\}$ rejects when Bob's
state is $P_j$. Then $q_{jj} = 0$, and $q_{kj} > 0$ for $k \neq j$, since
distinct pure states satisfy $\Tr[P_k P_j] < 1$. Let
\begin{equation}\label{eq:fc-delta}
  \delta = \min_{k \neq j} q_{kj} > 0
\end{equation}
be the smallest rejection probability of a wrong guess.
 
\paragraph{Memory and cyclic order.}
Alice's memory is $M = \mathbb{C}^L$ with orthonormal basis
$\ket{1}, \dots, \ket{L}$, one basis vector per label. Her persistent
register is $M \otimes A$ with $A = \mathbb{C}^d$, and Bob's register is
$B = \mathbb{C}^d$. The next label in the cyclic order is
\[
  k^+ = k + 1 \ \text{ for } k < L, \qquad L^+ = 1 .
\]
 
\paragraph{Alice's update channel.}
For each label $k$, define the two operators on $M \otimes A \otimes B$
\[
  Y_k = \ket{k}\bra{k} \otimes I_A \otimes P_k ,
  \qquad
  N_k = \ket{k}\bra{k} \otimes I_A \otimes (I_B - P_k).
\]
The operator $Y_k$ corresponds to ``the label is $k$ and the test says
yes'', and $N_k$ to ``the label is $k$ and the test says no''. Alice's
channel $\Phi : \Lin(M \otimes A \otimes B) \to \Lin(M \otimes A)$ is
\begin{equation}\label{eq:fc-channel}
  \Phi(X)
  = \sum_{k=1}^L \Tr[Y_k X]\; \ket{k}\bra{k} \otimes P_k^\top
  \;+\; \sum_{k=1}^L \Tr[N_k X]\; \ket{k^+}\bra{k^+} \otimes P_{k^+}^\top .
\end{equation}
In words: read the label $k$, test Bob's copy against $P_k$, keep the
label on ``yes'' and advance it on ``no'', and in either case prepare the
best response to the new label.

\begin{figure}[t]
  \centering
  \includegraphics[width=0.7\linewidth]{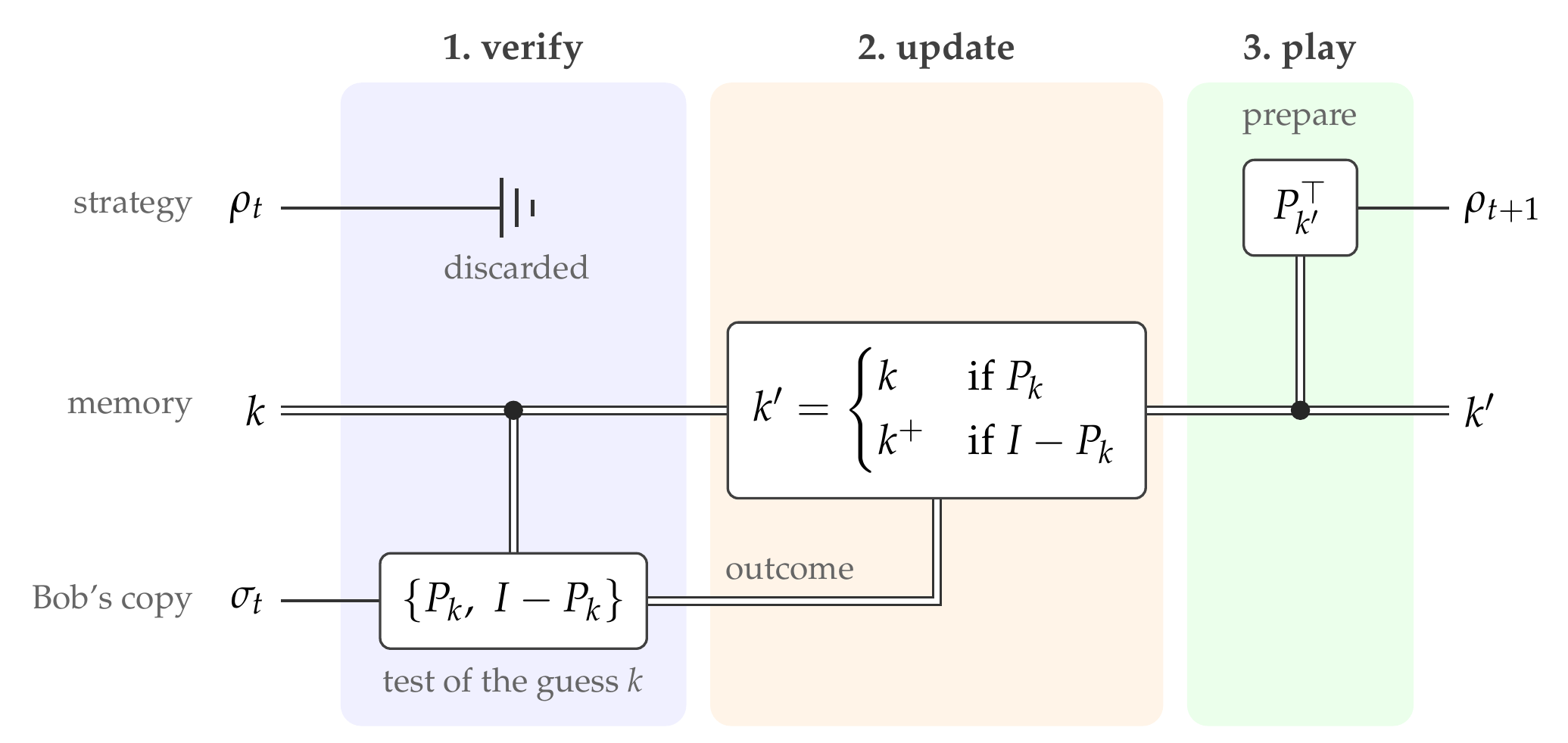}
  \caption{One round of Alice's update rule. Her previous strategy
  $\rho_t$ is discarded. The memory label $k$ selects the test
  $\{P_k, I - P_k\}$, which is applied to the received copy $\sigma_t$ of
  Bob's strategy. Outcome $P_k$ keeps the label, outcome $I - P_k$ advances
  it to $k^+$, and Alice prepares $P_{k'}^{\top}$ for the updated label $k'$.
  Double lines carry classical information.}
  \label{fig:alice-strategy}
\end{figure}

\begin{lemma}\label{lem:fc-channel}
The map $\Phi$ in~\eqref{eq:fc-channel} is a quantum channel. For every
input, its output has the form
$\sum_{k} c_k\, \ket{k}\bra{k} \otimes P_k^\top$ with $c_k \ge 0$ and
$\sum_k c_k$ equal to the trace of the input.
\end{lemma}
 
\begin{proof}
The operators $Y_k$ and $N_k$ are positive, and
$\sum_k (Y_k + N_k) = \sum_k \ket{k}\bra{k} \otimes I_A \otimes I_B = I$.
So $\{Y_k, N_k\}_k$ is a measurement, and $\Phi$ measures it and prepares
a fixed state for each outcome. Measure-and-prepare maps are channels.
The form of the output can be read off from~\eqref{eq:fc-channel}.
\end{proof}
 
Note that $\Phi$ ignores the off-diagonal part of the memory and
discards Alice's previous strategy. Its behavior therefore does not
depend on how the initial memory and strategy are correlated, and no
special initialization is needed.
 
\paragraph{The dynamics.}
Bob uses the replacement channel that prepares $P_j$, that is,
$\Psi^j_B(X) = \Tr[X]\, P_j$. Let $\omega_t \in \Dens(M \otimes A)$ be
Alice's joint state for memory and strategy and $\sigma_t$ be Bob's strategy. The
dynamics are
\begin{equation}\label{eq:fc-dynamics}
  \omega_{t+1} = \Phi(\omega_t \otimes \sigma_t), \qquad
  \sigma_{t+1} = P_j ,
\end{equation}
from an arbitrary initial pair $(\omega_0, \sigma_0)$. Alice's strategy is
the marginal $\rho_t = \Tr_M[\omega_t]$.
 
Both rules are autonomous. They are also uncoupled
(Definition~\ref{def:prelim-rule}): Alice's channel depends on the list
$P_1, \dots, P_L$, which is common to the whole family, but not on the
index $j$ of the game being played, while Bob's channel depends only on
his own target $P_j$.
 
\paragraph{Dynamics as a Markov chain.}
After the first round, the whole evolution can be described by a probability
vector over labels.
 
\begin{lemma}[Markov chain formulation]\label{lem:fc-markov}
In game $G_j$, for every $t \ge 1$,
\begin{equation}\label{eq:fc-form}
  \omega_t = \sum_{k=1}^L p_k(t)\; \ket{k}\bra{k} \otimes P_k^\top
\end{equation}
for a probability vector $p(t) = (p_1(t), \dots, p_L(t))$. Moreover,
$p(t)$ evolves as the distribution of a Markov chain $K_1, K_2, \dots$ on
the labels with transition probabilities
\begin{equation}\label{eq:fc-transitions}
  \Pr[K_{t+1} = k \mid K_t = k] = 1 - q_{kj}, \qquad
  \Pr[K_{t+1} = k^+ \mid K_t = k] = q_{kj}.
\end{equation}
\end{lemma}
 
\begin{proof}
By Lemma~\ref{lem:fc-channel}, every output of $\Phi$ has the
form~\eqref{eq:fc-form}, which proves the first claim. For $t \ge 1$,
Bob's state is $\sigma_t = P_j$, so the input to $\Phi$ is
$\omega_t \otimes P_j$. Using~\eqref{eq:fc-form} and
$\Tr[P_k^\top] = 1$,
\[
  \Tr[Y_k(\omega_t \otimes P_j)] = p_k(t)\, \Tr[P_k P_j]
    = p_k(t)\,(1 - q_{kj}),
  \qquad
  \Tr[N_k(\omega_t \otimes P_j)] = p_k(t)\, q_{kj}.
\]
Substituting into~\eqref{eq:fc-channel}: from label $k$, the weight
$p_k(t)(1 - q_{kj})$ stays at $k$ and the weight $p_k(t)\, q_{kj}$ moves to
$k^+$. These are the transitions~\eqref{eq:fc-transitions}.
\end{proof}
 
In the \emph{measurement realization} of the rule, Alice actually
records the outcomes. Then $K_t$ is her label at round $t$, and $p(t)$ is
its distribution. When $K_t = k$, her strategy at round $t$ is exactly
$P_k^\top$.

\subsection{Mixing time of the Markov chain and Nash convergence}\label{sec:finite-classical-theorem}

\begin{figure}[t]
  \centering
  \includegraphics[width=0.55\linewidth]{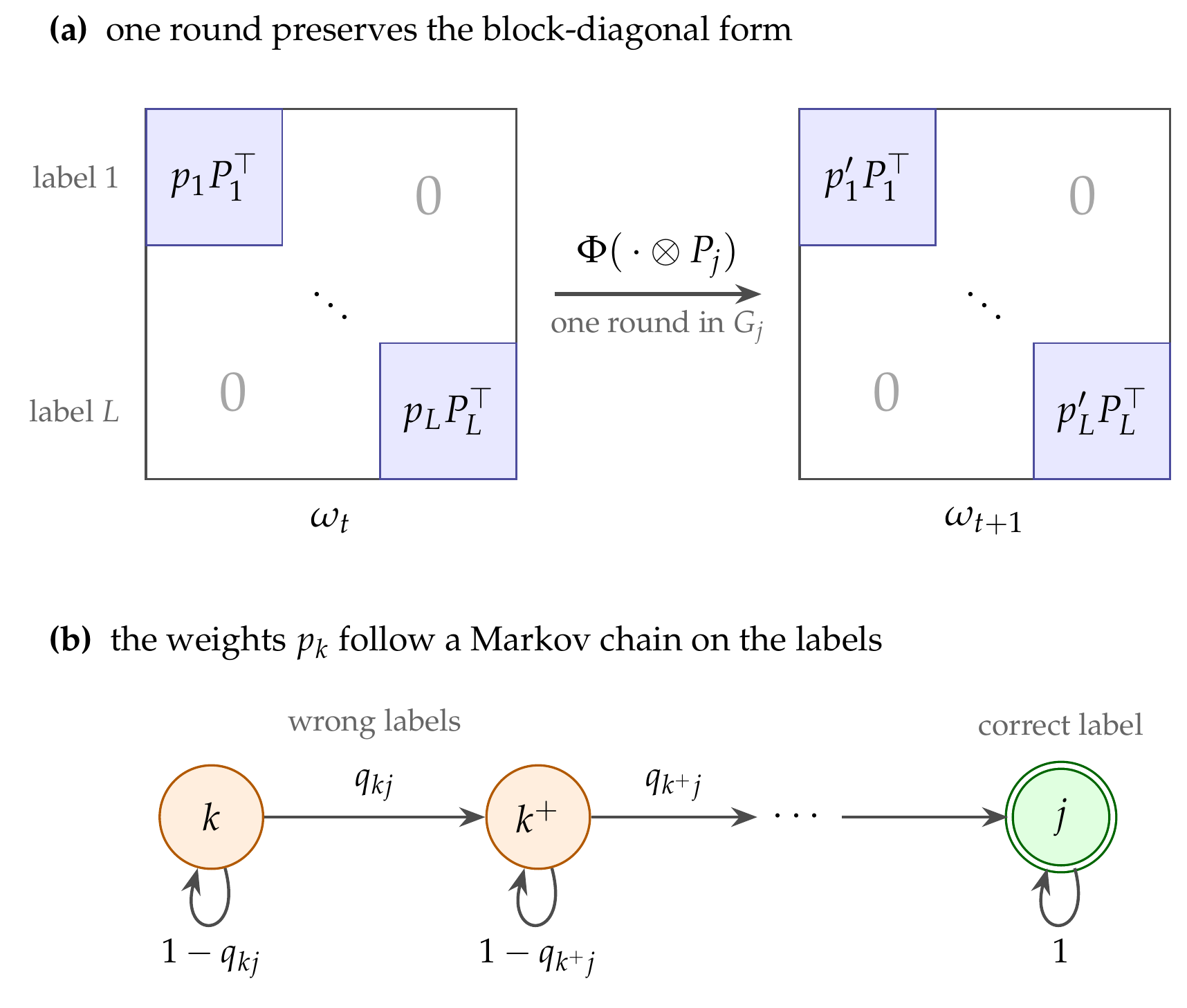}
  \caption{(a) After the first round, Alice's memory--strategy state is
  block diagonal in the label basis, with one block $p_k P_k^{\top}$ per
  label. One round of the dynamics in game $G_j$ returns a state of the
  same form and changes only the weights. (b) The weights therefore evolve
  as a Markov chain on the labels: a wrong label $k$ is kept with
  probability $1 - q_{kj}$ and proceed to $k^+$ with probability $q_{kj}$,
  and the correct label $j$ is never left.}
  \label{fig:label-markov}
\end{figure}

In this subsection, we bound the mixing time of the label chain and deduce
that the dynamics converge to Nash equilibrium. Fix a game $G_j$, and let
$Q$ be the transition matrix of the label chain $(K_t)_{t \ge 1}$ of
Lemma~\ref{lem:fc-markov}: from a label $k$, the chain stays at $k$
with probability $1 - q_{kj}$ and moves to $k^+$ with probability
$q_{kj}$. We index the chain by the number of transitions, so $Q^t(k, \cdot)$
is the distribution of the label $t$ rounds after the label was $k$. Let
$e_j$ denote the point mass on the correct label $j$. We measure the
distance to $e_j$ in total variation,
$\|\mu - \nu\|_{\mathrm{TV}} = \tfrac12 \sum_k |\mu_k - \nu_k|$, and define
the worst-case (total variation) distance after $t$ steps and the mixing time by
\begin{equation}\label{eq:fc-mix-def}
  d(t) = \max_{k} \bigl\| Q^t(k, \cdot) - e_j \bigr\|_{\mathrm{TV}}
  \qquad
  t_{\mathrm{mix}}(\varepsilon) = \min\{\, t \ge 0 : d(t) \le \varepsilon \,\}
  \quad (0 < \varepsilon < 1).
\end{equation}
For the label chain started at $k$, let
$T_j = \min\{\, t \ge 0 : K_t = j \,\}$ be the number of steps until it
first reaches $j$, and write $\Pr_k$ and $\mathbb{E}_k$ for probabilities
and expectations under such start.
 
\begin{theorem}[Mixing time and convergence]\label{thm:fc}
For the games~\eqref{eq:fc-games}, the autonomous uncoupled rule of
Section~\ref{sec:finite-classical-formal} uses an $L$-dimensional classical
memory for Alice and no memory for Bob. Fix a game $G_j$ and let $\delta$
be as in~\eqref{eq:fc-delta}.
\begin{enumerate}
  \item \textbf{Mixing time.} The label chain has the unique
        stationary distribution $e_j$. Its expected hitting time satisfies
        $\max_k \mathbb{E}_k[T_j] \le (L - 1)/\delta$, and its distance to
        stationarity satisfies
        \begin{equation}\label{eq:fc-dt}
          d(t) \le \exp\!\Bigl(-\frac{\delta t}{8}\Bigr)
          \qquad \text{for all } t \ge \frac{2(L-1)}{\delta}.
        \end{equation}
        Consequently, for every $0 < \varepsilon < 1$,
        \begin{equation}\label{eq:fc-tmix}
          L - 1 \;\le\; t_{\mathrm{mix}}(\varepsilon) \;\le\;
          \left\lceil \frac{2(L-1) + 8\ln(1/\varepsilon)}{\delta} \right\rceil .
        \end{equation}
  \item \textbf{Convergence of states.} Let
        $\Sigma_t = \omega_t \otimes \sigma_t$ and
        $\Sigma^* = \ket{j}\bra{j} \otimes P_j^\top \otimes P_j$. From every
        initial state $(\omega_0, \sigma_0)$ and for every $t \ge 0$,
        \begin{equation}\label{eq:fc-state-rate}
          \tfrac12 \|\Sigma_{1+t} - \Sigma^*\|_1 \le d(t).
        \end{equation}
        Hence after $1 + t_{\mathrm{mix}}(\varepsilon)$ rounds the joint
        state is within trace distance $\varepsilon$ of $\Sigma^*$, and
        the strategy profile $(\rho_t, \sigma_t)$ converges to the Nash
        equilibrium $(P_j^\top, P_j)$.
  \item \textbf{Finite convergence of trajectories.} In the
        measurement realization, the strategy profile reaches
        $(P_j^\top, P_j)$ after a finite number of rounds with probability
        one, and stays there forever.
\end{enumerate}
\end{theorem}
 
\begin{proof}
We work in game $G_j$ throughout. The proof proceeds in five steps.
 
\noindent\textbf{Step 1: total variation distance, hitting time, and
stationary distribution.} We first express the distance to $e_j$ in terms
of the hitting time $T_j$, then bound $T_j$, and finally use this bound to
identify the stationary distribution.

We first argue that the distance to $e_j$ is the probability of not having arrived. Since
$q_{jj} = 0$, once the chain reaches the correct label $j$ it stays there
forever. Hence, after $t$ steps, the chain is at $j$ if and only if it has
reached $j$ within those $t$ steps, that is,
$Q^t(k, j) = \Pr_k[T_j \le t]$. Because $e_j$ puts all of its mass on $j$,
the total variation distance between $Q^t(k, \cdot)$ and $e_j$ is the mass
that $Q^t(k, \cdot)$ puts on wrong labels:
\begin{equation}\label{eq:fc-tv-survival}
  \bigl\| Q^t(k, \cdot) - e_j \bigr\|_{\mathrm{TV}}
  = \tfrac12 \Bigl( \sum_{k' \neq j} Q^t(k, k') + 1 - Q^t(k, j) \Bigr)
  = 1 - Q^t(k, j)
  = \Pr\nolimits_k[T_j > t].
\end{equation}
Taking the maximum over the starting label gives
$d(t) = \max_k \Pr_k[T_j > t]$. In words, mixing of the label chain is
the same as arrival at the correct label.

Next, we compute the expected hitting time. At a wrong label $k \neq j$, the chain
leaves $k$ in each round with probability $q_{kj} \ge \delta$, and when it
leaves, it moves to the next label $k^+$. The number of rounds spent at
$k$, counting the round in which it leaves, is therefore geometric with
mean $1/q_{kj} \le 1/\delta$. Following the cyclic order from any starting
label, the chain passes through at most $L - 1$ wrong labels before
reaching $j$. Adding the expected waiting times gives
$\mathbb{E}_k[T_j] \le (L - 1)/\delta$ for every start $k$. In particular,
$T_j$ is finite with probability one, so $\Pr_k[T_j > t] \to 0$ as
$t \to \infty$ for each $k$. Since there are finitely many labels, this
gives $d(t) \to 0$.

Finally, we identify the stationary distribution. The point mass $e_j$ is stationary because the chain never leaves $j$. Conversely, let $\pi$ be any
stationary distribution, so that $\pi Q^t = \pi$ for every $t$. The mass
that $\pi$ assigns to wrong labels then equals the mass that $\pi Q^t$
assigns to them, which by~\eqref{eq:fc-tv-survival} is
\[
  \sum_k \pi_k \Pr\nolimits_k[T_j > t] \le d(t).
\]
Letting $t \to \infty$ shows that $\pi$ puts no mass on wrong labels, so
$\pi = e_j$.
 
\noindent\textbf{Step 2: exponential decay of $d(t)$.} We realize the
chain with independent coin flips. Let $U_1, U_2, \dots$ be independent
and uniform on $[0, 1]$. If the label after $s - 1$ steps is $k$, the
chain moves to $k^+$ at step $s$ if $U_s < q_{kj}$, and stays at $k$
otherwise. This has exactly the transition probabilities of $Q$, so it
has the same distribution as the label chain. (Independence across rounds
reflects the fact that Alice tests a fresh, independent copy of Bob's state
in every round.)
 
Call step $s$ a \emph{lucky step} if $U_s < \delta$. Because
$q_{kj} \ge \delta$ for every wrong label $k$, a lucky step taken at a wrong
label always advances the chain. Let $X_t$ be the number of lucky steps
among the first $t$ steps. Then $X_t$ has the binomial distribution with
$t$ trials and success probability $\delta$, whatever the starting label.
Now suppose $T_j > t$. Then the chain is at a wrong label during each of
the first $t$ steps, so each of the $X_t$ lucky steps is an advance. But
the chain reaches $j$ after at most $L - 1$ advances, so it can make at most
$L - 2$ advances without arriving. Hence
\[
  \{T_j > t\} \subseteq \{X_t \le L - 2\}.
\]
Let $\mu = \mathbb{E}[X_t] = \delta t$. The (multiplicative) Chernoff bound
for binomial variables gives
\[\Pr[X_t \le \mu/2] \le e^{-\mu/8}.\] If $t \ge 2(L - 1)/\delta$, then
$L - 2 < \mu/2$, and therefore, for every start $k$,
\[
  \Pr\nolimits_k[T_j > t]
  \le \Pr[X_t \le \mu/2]
  \le \exp\!\Bigl(-\frac{\delta t}{8}\Bigr).
\]
Taking the maximum over $k$ and using Step~1 proves~\eqref{eq:fc-dt}.
 
\noindent\textbf{Step 3: the mixing time.} For the upper bound
in~\eqref{eq:fc-tmix}, let
$t_0 = \lceil (2(L - 1) + 8\ln(1/\varepsilon))/\delta \rceil$. Then
$t_0 \ge 2(L - 1)/\delta$, so~\eqref{eq:fc-dt} applies and gives
$d(t_0) \le \exp(-\delta t_0/8) \le \varepsilon$. For the lower bound,
start the chain at $j^+$, the label that follows $j$ in the cyclic order.
The chain moves by at most one label per step and must make $L - 1$
advances to reach $j$, so $\Pr_{j^+}[T_j > t] = 1$ for every $t < L - 1$.
Hence $d(t) = 1 > \varepsilon$ for $t < L - 1$, and
$t_{\mathrm{mix}}(\varepsilon) \ge L - 1$. This proves part~1.
 
\noindent\textbf{Step 4: the convergence of states.} By
Lemma~\ref{lem:fc-markov}, for $t \ge 1$ we have
$\omega_t = \sum_k p_k(t)\, \ket{k}\bra{k} \otimes P_k^\top$ and
$\sigma_t = P_j$. Therefore
\[
  \Sigma_t - \Sigma^*
  = \left( \sum_{k \neq j} p_k(t)\, \ket{k}\bra{k} \otimes P_k^\top
     \;-\; \bigl(1 - p_j(t)\bigr)\, \ket{j}\bra{j} \otimes P_j^\top \right)
     \otimes P_j .
\]
The terms act on mutually orthogonal memory blocks, so the trace norm is
the sum of the trace norms of the terms:
\begin{equation}\label{eq:fc-distance}
  \tfrac12 \|\Sigma_t - \Sigma^*\|_1
  = \tfrac12 \left( \sum_{k \neq j} p_k(t) + 1 - p_j(t) \right)
  = 1 - p_j(t)
  = \bigl\| p(t) - e_j \bigr\|_{\mathrm{TV}} .
\end{equation}
The trace distance of the quantum state to its target thus equals the total variation distance of the label distribution to $e_j$. The label distribution after round $1$ is $p(1)$, and $t$ steps later it is $p(1+t) = \sum_k p_k(1)\, Q^t(k, \cdot)$. By convexity of the total variation distance, we get
\[
  \tfrac12 \|\Sigma_{1+t} - \Sigma^*\|_1
  = \bigl\| p(1+t) - e_j \bigr\|_{\mathrm{TV}}
  \le \sum_k p_k(1)\, \bigl\| Q^t(k, \cdot) - e_j \bigr\|_{\mathrm{TV}}
  \le d(t),
\]
which is~\eqref{eq:fc-state-rate}. Since $d(t) \to 0$ and partial traces
do not increase the trace norm, $\rho_t \to P_j^\top$ and
$\sigma_t \to P_j$. This proves part~2.
 
\noindent\textbf{Step 5: finite-time convergence of trajectories.} In the
measurement realization, once the label reaches $j$, Alice prepares
$P_j^\top$ and keeps the label $j$ forever, while Bob plays $P_j$ from
round $1$ on. Since $T_j$ is finite with probability one by Step~1, the
profile equals $(P_j^\top, P_j)$ after finitely many rounds and stays
there. This proves part~3.
\end{proof}

\begin{corollary}[The families of Section~\ref{sec:memoryless-impossibilty}]
\label{cor:fc-families}
The four games of Section~\ref{sec:memoryless-games} are solved by the
rule above with $L = 4$ labels, that is, two classical bits, and
$\max_k \mathbb{E}_k[T_j] \le 6$. The $4^n$ games of
Section~\ref{sec:memoryless-nqubit} are solved with $L = 4^n$ labels,
that is, $2n$ classical bits, and $\max_k \mathbb{E}_k[T_j] \le 2(4^n - 1)$.
\end{corollary}
 
\begin{proof}
Two distinct states of $\mathcal{S}$ have squared overlap $0$ (for the
pair $\ket{0}, \ket{1}$) or $\tfrac12$ (for every other pair). The squared
overlap of two product states is the product of the squared overlaps of
their factors. Hence two distinct states in $\mathcal{S}_n$ have squared
overlap at most $\tfrac12$, with equality when they differ in exactly one
factor and that factor pair is not $\{\ket{0}, \ket{1}\}$. Therefore
$\delta = \tfrac12$ for both families, and Theorem~\ref{thm:fc} gives the
stated bounds.
\end{proof}
 
\begin{remark}[What the memory does, and what it costs]
\label{rem:fc-discussion}
The memory makes the rigidity constraints harmless. The equilibrium
persistent states $\ket{j}\bra{j} \otimes P_j^\top$ are mutually orthogonal
for different $j$, so the inputs constrained by the fixed-point conditions
are orthogonal, and the overlap argument of Lemma~\ref{lem:four-state}
yields nothing. Moreover, these inputs span only a small subspace of the
input space $M \otimes A \otimes B$, so they do not determine the channel.
The price is that convergence is not monotone. While the label is wrong,
Alice plays the best response to the wrong target, so her strategy may move
\emph{away} from equilibrium before the label settles, even if she started
at the equilibrium strategy. The object that converges is the full
memory--strategy state, as in part~2 of Theorem~\ref{thm:fc}. Finally, the
construction relies on the family being finite and known, since Alice must
store one label per possible target. It says nothing about families with
infinitely many targets, such as the family of all pure states;
Section~\ref{sec:barrier} shows that this limitation is unavoidable.
\end{remark}

\section{An Impossibility for Finite Quantum Memory}\label{sec:barrier}
 
Section~\ref{sec:finite-classical} showed that a few classical memory bits
let Alice reach equilibrium in any \emph{finite} family of conjugation
games. In this section we show that this success cannot be extended to the
\emph{continuum} of all pure targets: no learner with a fixed
finite-dimensional memory, classical or quantum, learns to transpose
every pure qubit-state, even against an opponent who plays the same state
in every round. Section~\ref{sec:fm-overview} sets up the model, states the
main theorem, and explains the ideas behind its proof.
Sections~\ref{sec:fm-exp},~\ref{sec:fm-unot}, and~\ref{sec:fm-combine}
carry out the three parts of the proof, and the end of
Section~\ref{sec:fm-combine} derives the consequence for uncoupled game
dynamics.
 
\subsection{Overview and main ideas}\label{sec:fm-overview}
 
\paragraph{The model.}
Alice's \emph{persistent register} is a finite-dimensional Hilbert space
$H$. It contains her strategy together with all of her memory, classical
or quantum. In every round she receives $b$ fresh copies of Bob's current
strategy, where $b \ge 1$ is fixed, and updates her register by a fixed
channel
\[
  \mathcal{A} : \Lin\bigl(H \otimes (\mathbb{C}^2)^{\otimes b}\bigr) \to \Lin(H).
\]
Her current strategy is extracted from the register by a fixed
\emph{readout} channel $R : \Lin(H) \to \Lin(\mathbb{C}^2)$; when the
strategy is a subsystem of $H$, the readout $R$ is a partial trace. Alice starts from an
initial state $\Psi_0 \in \Dens(H)$ that does not depend on Bob.
 
We study the simplest possible opponent: Bob plays a fixed pure qubit-state
$P$ in every round. Since the same channel and the same input are used in
every round, Alice's register evolves by powers of a single linear map,
\begin{equation}\label{eq:fm-FP}
  F_P(X) = \mathcal{A}\bigl(X \otimes P^{\otimes b}\bigr),
  \qquad
  \Psi_t(P) = F_P^{\,t}(\Psi_0),
  \qquad
  \Sigma_t(P) = R\bigl(\Psi_t(P)\bigr).
\end{equation}
Alice's best response to $P$ is $P^\top$
(Lemma~\ref{lem:prelim-conj-ne}), so we measure her error by the
\emph{infidelity}
\begin{equation}\label{eq:fm-error}
  f_t(P) = 1 - \Tr\bigl[P^\top \Sigma_t(P)\bigr] \in [0, 1].
\end{equation}
Since $P^\top$ is pure, $f_t(P) \to 0$ if and only if $\Sigma_t(P) \to P^\top$.
 
\begin{theorem}[Finite-memory barrier]\label{thm:fm-barrier}
For every finite-dimensional $H$, every fixed $b \ge 1$, channels
$\mathcal{A}$ and $R$, and every initial state
$\Psi_0$, there is a pure state $P$ such that $\Sigma_t(P)$ does not
converge to $P^\top$.
\end{theorem}
 
In words, whatever finite memory and update rule Alice uses, there is a
stationary opponent against which she never learns her best response. The
proof plays two facts against each other: finite-memory dynamics can only
converge exponentially fast (Ideas~1 and~2 below), whereas quantum
mechanics allows transposition only with polynomially small error
(Idea~3).

\paragraph{Idea 1: finite-memory dynamics learn exponentially fast or not
at all.}
For a fixed $P$, the error $f_t(P)$ is a linear function of
$F_P^{\,t}(\Psi_0)$, which is the $t$-th power of one linear map on the
finite-dimensional space $\Lin(H)$. Such sequences are very rigid. We will
show that if $f_t(P) \to 0$, then it must converge exponentially fast.
The underlying principle comes from dynamical systems.
 
Conley's fundamental theorem of dynamical systems~\cite{Conley1978}
says that every continuous dynamical system on a compact metric space
splits into two parts: a \emph{chain-recurrent} part, on which orbits keep
returning close to where they have been, and a \emph{gradient-like} part,
on which a continuous Lyapunov function strictly decreases along orbits.

For linear maps with bounded powers, such a split is a classical theorem of
operator theory—the Jacobs–de Leeuw–Glicksberg (JdLG)
decomposition~\cite{Jacobs1956,Jacobs1957,deLeeuwGlicksberg1961}. In its
general form, it concerns a bounded operator $T$ on a Banach space whose
orbits are relatively weakly compact. It splits the space into a
\emph{reversible} part, the closed linear span of the eigenvectors of $T$
with eigenvalues of modulus one, and a \emph{stable} part, formed by the
vectors whose orbit has $0$ in its weak closure. In our setting $\Lin(H)$ is
finite dimensional and $F_P$ is power-bounded, so the hypothesis holds
automatically and the decomposition takes an explicit form. The space
$\Lin(H)$ decomposes into a \emph{recurrent subspace}, on which $F_P$ acts
like a rotation on a torus, and a \emph{transient subspace}, on which
$F_P^{\,t}$ converges geometrically. 

For quantum channels these are the
attractor (or peripheral) subspace and its
complement~\cite{AmatoFacchiKonderak2023}. Recurrence forbids the recurrent
part from contributing to an error that tends to zero because a recurrent
orbit keeps coming back to its starting configuration. Thus, only the
transient part remains, and it converges exponentially.
 
Conley's theorem itself is topological, so it does not give convergence rates. For instance, the maps $x \mapsto x/2$ and $x \mapsto x - x^3$ on
$[0, \tfrac12]$ are topologically conjugate near $0$, yet their orbits
converge like $2^{-t}$ and like $t^{-1/2}$, respectively. The exponential rate comes from the extra structure of our problem, namely
linearity and finite dimension, which is exactly what the JdLG
decomposition exploits. To keep the paper self-contained,
Section~\ref{sec:fm-exp} proves its finite-dimensional case directly from
the Jordan form, with an explicit rate.
 
\paragraph{Idea 2: from a rate for each state to a rate on a patch.}
Idea~1 gives constants $C_P < \infty$ and $q_P < 1$ with
$f_t(P) \le C_P\, q_P^t$, but these constants may depend badly on $P$. The
Baire category theorem states that a compact metric space is not a
countable union of closed sets each of which contains no open set. We
group the states $P$ according to how good their constants are. This gives
countably many closed sets that cover the Bloch sphere. By Baire's theorem, one of
them must contain an open patch $U$. Hence there are \emph{uniform} constants
$C$ and $q < 1$ with $f_t(P) \le C q^t$ for every $P \in U$.
 
\paragraph{Idea 3: quantum mechanics forbids fast transposition.}
The universal-NOT (U-NOT) is the ideal operation that maps every
pure qubit-state $P$ to its orthogonal complement $I - P$, that is, the
state with the opposite Bloch vector. It is not a quantum channel: on the
Bloch sphere, it is the inversion $r \mapsto -r$, which reverses
the orientation and is therefore implemented by an antiunitary operator.
Bu\v{z}ek, Hillery, and Werner~\cite{BuzekHilleryWerner1999} found the best
physical approximation: given $N$ copies of an unknown pure state, every
channel produces the orthogonal state with average fidelity at most
$\frac{N+1}{N+2}$, and this value is attained by first estimating the state
and then preparing the state orthogonal to the estimate.
 
Transposition and the U-NOT are equivalent up to a fixed unitary. As noted
in the preliminaries, transposition reflects the Bloch sphere through the
$x$--$z$ plane, sending $r = (r_1, r_2, r_3)$ to $(r_1, -r_2, r_3)$.
Conjugation by the Pauli matrix $Y$ rotates by $\pi$ on the $y$-axis,
sending $(r_1, r_2, r_3)$ to $(-r_1, r_2, -r_3)$. Composing those two will send
$r$ to $-r$, so 
\begin{equation}\label{eq:fm-unot-transpose}
  Y P^\top Y = I - P \qquad \text{for every pure state } P.
\end{equation}
Consequently, transposing $N$ copies of an unknown pure state also has
average infidelity at least $\frac{1}{N+2}$. This error can decrease only
\emph{polynomially} in the number of copies.
 
\paragraph{Combining the ideas.}
Suppose, for contradiction, that Alice's learner converges for every $P$.
Ideas~1 and~2 give a patch $U$ on which it converges uniformly and
exponentially fast. We turn the learner into a transposition device that
works for \emph{every} pure state. The device uses a few copies to estimate
the unknown state roughly, applies one of finitely many fixed rotations
that moves the state into the patch $U$, runs the learner for $m$ rounds on
rotated copies, and finally undoes the rotation on the output. It consumes
$N = (b + 3)m$ copies and has infidelity exponentially small in $m$. This
contradicts the bound $\frac{1}{N+2}$ of Idea~3, which decreases only like
$1/m$. Hence the learner cannot converge for every $P$.
 
The remaining subsections make each step precise: exponential convergence
(Section~\ref{sec:fm-exp}), the U-NOT bound (Section~\ref{sec:fm-unot}),
and the combination (Section~\ref{sec:fm-combine}).
 
\subsection{Finite-memory dynamics converges exponentially fast, if at all}
\label{sec:fm-exp}
 
Throughout this subsection, $V$ is a finite-dimensional complex vector
space with a fixed norm, and $T : V \to V$ is linear. We say that $T$ is
\emph{power-bounded} if $\sup_{t \ge 0} \|T^t\| < \infty$, where $\|\cdot\|$
is the induced operator norm. In finite dimension all norms are equivalent,
so this property does not depend on the chosen norm.
 
\begin{lemma}[Channels are power-bounded]\label{lem:fm-power-bounded}
Every quantum channel $F : \Lin(H) \to \Lin(H)$ satisfies
$\|F^t(X)\|_1 \le 2\|X\|_1$ for all $X \in \Lin(H)$ and all
$t \ge 0$. In particular, $F$ is power-bounded.
\end{lemma}
 
\begin{proof}
First let $A$ be Hermitian, and let $A = A_+ - A_-$ with
$A_\pm \succeq 0$ and $\|A\|_1 = \Tr[A_+] + \Tr[A_-]$. Since
$F^t$ is positive and trace preserving,
\[
  \|F^t(A)\|_1
  \le \|F^t(A_+)\|_1 + \|F^t(A_-)\|_1
  = \Tr[A_+] + \Tr[A_-] = \|A\|_1 .
\]
A general $X$ can be written as $X = A + iB$ with Hermitian
$A = (X + X^\dagger)/2$ and $B = (X - X^\dagger)/(2i)$, and
$\|A\|_1, \|B\|_1 \le \|X\|_1$. The triangle inequality then gives
$\|F^t(X)\|_1 \le \|A\|_1 + \|B\|_1 \le 2\|X\|_1$.
\end{proof}
 
The next lemma is the finite-dimensional case of the JdLG
decomposition. In its terminology, $V_{\mathrm{rec}}$ is the "reversible part" and $V_{\mathrm{trans}}$
is the "stable part." We include a short proof using the Jordan form, which also gives the geometric rate in part~2. Also, see \cite[Ch.~6]{Wolf2012} for related discussion on the spectral properties of quantum channels and \cite[Fact 2.4 and Theorem 2.12]{Bauschke2014convergencematrices} for a similar argument on matrices.

\begin{lemma}[Recurrent and transient parts]\label{lem:fm-split}
Let $T$ be power-bounded. Then $V = V_{\mathrm{rec}} \oplus V_{\mathrm{trans}}\,$ 
for two $T$-invariant subspaces with the following properties.
\begin{enumerate}
  \item \textbf{Recurrent part.} $V_{\mathrm{rec}}$ has a basis of
        eigenvectors of $T$ whose eigenvalues all have modulus one.
  \item \textbf{Transient part.} There are constants $C < \infty$ and
        $0 < q < 1$ such that $\|T^t x\| \le C q^t \|x\|$ for all
        $x \in V_{\mathrm{trans}}$ and all $t \ge 0$.
\end{enumerate}
\end{lemma}
 
\begin{proof}
By the Jordan canonical form theorem~\cite[Sec.~3.1]{HornJohnson2012},
$V$ is the direct sum of the generalized eigenspaces
$V_\lambda = \ker (T - \lambda I)^{\dim V}$ of $T$, each of which is
$T$-invariant.

We first show that every eigenvalue of $T$ lies in the closed unit disk.
If $Tv = \lambda v$ with $v \neq 0$, then $\|T^t v\| = |\lambda|^t \|v\|$,
and power-boundedness forces $|\lambda| \le 1$.

Next, we show that eigenvalues on the unit circle have no nontrivial
Jordan blocks. Let $|\lambda| = 1$, and write $T = \lambda(I + N)$ on
$V_\lambda$, where $N$ is nilpotent. If $N \neq 0$, choose
$v \in V_\lambda$ with $Nv \neq 0$ and $N^2 v = 0$. Then
$T^t v = \lambda^t (v + tNv)$, whose norm grows linearly in $t$,
contradicting power-boundedness. Hence $N = 0$, so $T = \lambda I$ on
$V_\lambda$, and every nonzero vector of $V_\lambda$ is an eigenvector.

Let $V_{\mathrm{rec}}$ be the sum of the $V_\lambda$ with $|\lambda| = 1$,
and $V_{\mathrm{trans}}$ the sum of those with $|\lambda| < 1$. Part~1
follows from the previous paragraph. For part~2, note that on each $V_\lambda$, the map $T - \lambda I$ is nilpotent (by definition),
so the only eigenvalue of $T|_{V_\lambda}$ is $\lambda$. Since
$V_{\mathrm{trans}}$ is the direct sum of the $T$-invariant spaces $V_\lambda$
with $|\lambda| < 1$, every eigenvalue of $T|_{V_{\mathrm{trans}}}$ is one of
these finitely many $\lambda$. Hence the spectral radius
$r = \max\{|\lambda| : V_\lambda \subseteq V_{\mathrm{trans}}\}$ of
$T|_{V_{\mathrm{trans}}}$ satisfies $r < 1$. If
$V_{\mathrm{trans}} = \{0\}$ there is nothing to prove. Otherwise,
Gelfand's formula (\cite[Cor.~5.6.14]{HornJohnson2012})
\[\lim_{t \to \infty} \|T^t|_{V_{\mathrm{trans}}}\|^{1/t} = r\]
shows that for any $q \in (r, 1)$ we
have $\|T^t|_{V_{\mathrm{trans}}}\| \le q^t$ for all $t \ge t_0$, for some
$t_0$. Part~2 follows with
$C = \max\bigl\{1, \max_{t < t_0} \|T^t|_{V_{\mathrm{trans}}}\|\, q^{-t}\bigr\}$.
\end{proof}
 
The recurrent part really admits the recurrent behavior. Its orbits return arbitrarily close
to their starting point at arbitrarily late times.

\begin{figure}[t]
\centering
\includegraphics[width=0.52\linewidth]{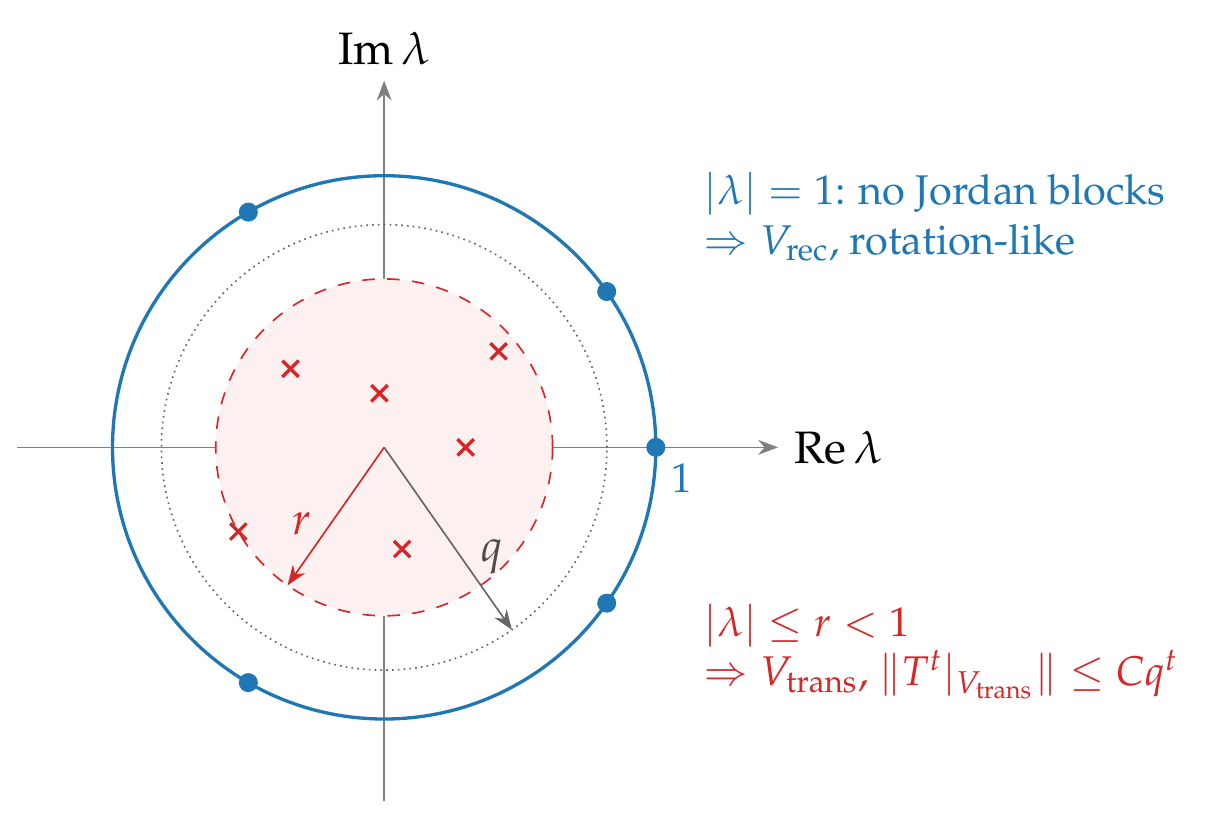}
\caption{Spectrum of a power-bounded operator $T$. Power-boundedness confines
all eigenvalues to the closed unit disk and forbids non-trivial Jordan blocks on the unit
circle. Eigenvalues on the circle (dots) span the recurrent part
$V_{\mathrm{rec}}$. The remaining finitely many eigenvalues (crosses) lie in the
disk of radius $r < 1$, and every $q \in (r,1)$ gives a geometric rate on the
transient part $V_{\mathrm{trans}}$.}
\label{fig:fm-spectrum}
\end{figure}
 
\begin{lemma}[Recurrence]\label{lem:fm-recurrence}
Let $T$ be power-bounded and $x \in V_{\mathrm{rec}}$. Then there are
times $s_1 < s_2 < \cdots$ with $s_n \to \infty$ and $T^{s_n} x \to x$.
\end{lemma}
 
\begin{proof}
By Lemma~\ref{lem:fm-split}, write $x = \sum_{i=1}^m x_i$, where $x_i$ is
an eigenvector of $T$ with eigenvalue $\lambda_i$ and $|\lambda_i| = 1$.
Then $T^s x = \sum_i \lambda_i^s x_i$, so it suffices to find, for every
$\varepsilon > 0$ and every $S$, a time $s \ge S$ with
$|\lambda_i^s - 1| < \varepsilon$ for all $i$.
 
The points $z_t = (\lambda_1^t, \dots, \lambda_m^t)$ lie on the torus
$\{z \in \mathbb{C}^m : |z_i| = 1\}$, which is compact. Cover it by finitely
many sets of diameter less than $\varepsilon$ in the maximum norm, say $K$
of them. Among the $K + 1$ points $z_0, z_S, z_{2S}, \dots, z_{KS}$, two lie
in the same set, say $z_{aS}$ and $z_{bS}$ with $a < b$. Then
$|\lambda_i^{bS} - \lambda_i^{aS}| < \varepsilon$ for every $i$. Dividing
by $\lambda_i^{aS}$, which has modulus one, gives
$|\lambda_i^{(b-a)S} - 1| < \varepsilon$. Therefore, the time $s = (b - a)S \ge S$
has the required property.
\end{proof}
We can now prove the main result of this subsection.
 
\begin{lemma}[Convergence implies exponential convergence]
\label{lem:fm-exp}
Let $T$ be power-bounded, $x \in V$, and $\varphi : V \to \mathbb{C}$ be linear.
If $\varphi(T^t x) \to 0$ as $t \to \infty$, then there are constants
$C < \infty$ and $0 < q < 1$ with
\[
  |\varphi(T^t x)| \le C q^t \qquad \text{for all } t \ge 0.
\]
\end{lemma}
 
\begin{proof}
Decompose $x = x_{\mathrm{rec}} + x_{\mathrm{trans}}$ according to
Lemma~\ref{lem:fm-split}, and set
$a_t = \varphi(T^t x_{\mathrm{rec}})$ and $b_t = \varphi(T^t x_{\mathrm{trans}})$.
By part~2 of Lemma~\ref{lem:fm-split},
$|b_t| \le \|\varphi\|\, C' q^t \|x_{\mathrm{trans}}\|$, which tends to $0$.
Since $a_t + b_t = \varphi(T^t x) \to 0$, also $a_t \to 0$. We claim that $a_t = 0$ for every $t$. Fix $t$, and let $s_n \to \infty$ be the return times of Lemma~\ref{lem:fm-recurrence} for $x_{\mathrm{rec}}$.
Since $T^t$ and $\varphi$ are linear maps on a finite-dimensional space, they
are continuous, and so
\[
  a_{t + s_n} = \varphi\bigl(T^t\, T^{s_n} x_{\mathrm{rec}}\bigr)
  \;\longrightarrow\; \varphi\bigl(T^t x_{\mathrm{rec}}\bigr) = a_t .
\]
The left-hand side tends to $0$ because $a_t \to 0$. Hence $a_t = 0$. Therefore $\varphi(T^t x) = b_t$ for every $t$, and the bound on $b_t$ proves
the lemma with $C = C'\|\varphi\|\, \|x_{\mathrm{trans}}\|$, or with any $C > 0$
if this quantity vanishes.
\end{proof}
 
In words, a recurrent orbit keeps revisiting its starting configuration,
so its contribution to the error cannot fade unless it was zero all along.
Everything that does fade lies in the transient part, which fades
exponentially.

\begin{figure}[t]
\centering
\includegraphics[width=0.8\linewidth]{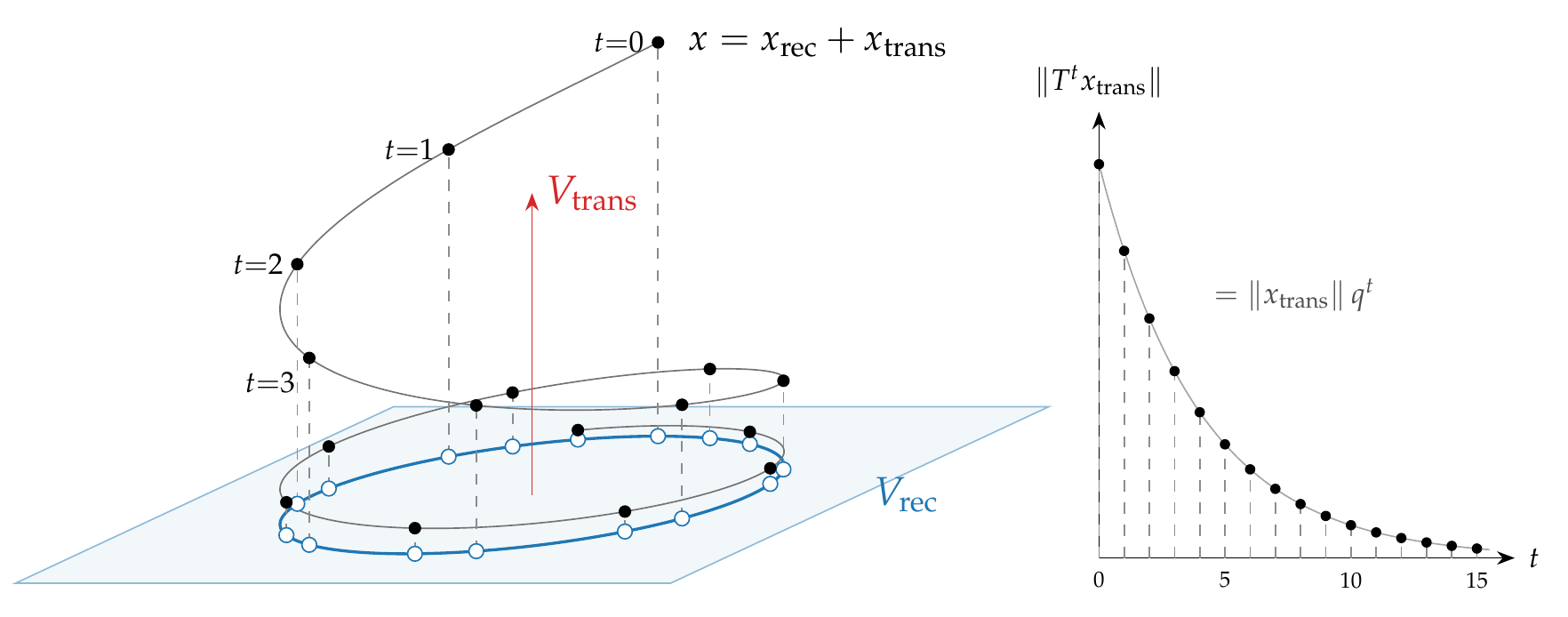}
\caption{An orbit $T^t x$ for $T$ acting as an irrational rotation on
$V_{\mathrm{rec}}$ (horizontal plane) and as multiplication by $q<1$ on
$V_{\mathrm{trans}}$ (vertical axis). The \emph{height} of each orbit
point above the plane (dashed gray line) is its transient component
$T^t x_{\mathrm{trans}}$. The panel on the right plots this height against
$t$; it shrinks geometrically, as $\|x_{\mathrm{trans}}\|\,q^t$. Hence the
orbit spirals down onto the circle. The \emph{shadow} of each orbit point on
the plane (open circle) is its recurrent component $T^t x_{\mathrm{rec}}$.
The shadows stay on the circle and never settle down: they keep returning
close to $x_{\mathrm{rec}}$ (Lemma~\ref{lem:fm-recurrence}).}
\label{fig:fm-orbit}
\end{figure}
 
We apply Lemma~\ref{lem:fm-exp} to Alice's learner. For a fixed pure
state $P$, take $V = \Lin(H)$, $T = F_P$, $x = \Psi_0$, and the
linear functional $\varphi(X) = \Tr[(I - P^\top)R(X)]$. Then
$\varphi(T^t x) = f_t(P)$ by~\eqref{eq:fm-error}, and $F_P$ is
power-bounded by Lemma~\ref{lem:fm-power-bounded}. Hence
\begin{equation}\label{eq:fm-pointwise}
  f_t(P) \to 0
  \quad\Longrightarrow\quad
  f_t(P) \le C_P\, q_P^{\,t} \ \text{ for all } t,
  \text{ for some } C_P < \infty,\ q_P < 1 .
\end{equation}
The constants depend on $P$. The final lemma of this subsection makes them
uniform on an open set of states.
 
\begin{lemma}[Uniform exponential convergence on a patch]
\label{lem:fm-patch}
Suppose that $f_t(P) \to 0$ for every pure qubit-state $P$. Then there are
a nonempty open set $U$ of pure states and constants $C < \infty$ and
$0 < q < 1$ such that
\[
  f_t(P) \le C q^t \qquad \text{for all } P \in U \text{ and all } t \ge 0.
\]
\end{lemma}
 
\begin{proof}
Identify the pure states with the Bloch sphere $S^2$, which is a compact
metric space. For each fixed $t$, the map $P \mapsto f_t(P)$ is continuous,
because $F_P$ depends polynomially on the entries of $P$ through
$P^{\otimes b}$, and the composition and trace are continuous. For integers
$j \ge 1$ and $k \ge 2$, let
\[
  E_{j,k} = \bigl\{ P : f_t(P) \le j\,(1 - 1/k)^t \text{ for every integer } t \ge 0 \bigr\}.
\]
Each $E_{j,k}$ is an intersection of closed sets and is therefore closed.
By~\eqref{eq:fm-pointwise}, every $P$ belongs to some $E_{j,k}$ by choosing
$j \ge C_P$ and $k$ with $1 - 1/k \ge q_P$. Thus the countably many closed
sets $E_{j,k}$ cover $S^2$. By the Baire category theorem, some $E_{j,k}$
contains a nonempty open set $U$. The lemma holds with $C = j$ and
$q = 1 - 1/k$.
\end{proof}
 
\subsection{The universal-NOT bound and the transposition bound}\label{sec:fm-unot}

We now make precise the quantum limit behind Idea~3 of
Section~\ref{sec:fm-overview}. Throughout, integrals $\int \cdot\, dP$ over
pure qubit-states are taken with respect to the unitarily invariant
probability measure, that is, the uniform measure on the Bloch sphere.

Our starting point is the optimal universal-NOT bound by Bu\v{z}ek, Hillery, and
Werner~\cite{BuzekHilleryWerner1999}. Given $N$ copies of an unknown pure
state $P$, the ideal output $I - P$ cannot be produced by any channel, and
the best achievable fidelity coincides with that of optimal state
estimation.

\begin{theorem}[Optimal universal-NOT~\cite{BuzekHilleryWerner1999}]
\label{thm:fm-unot}
For every $N \ge 1$ and every channel
$\Lambda : \Lin\bigl((\mathbb{C}^2)^{\otimes N}\bigr) \to \Lin(\mathbb{C}^2)$,
\[
  \int \Tr\bigl[(I - P)\, \Lambda(P^{\otimes N})\bigr]\, dP
  \;\le\; \frac{N + 1}{N + 2}.
\]
The bound is attained by estimating $P$ from the $N$ copies and preparing
the state orthogonal to the estimate.
\end{theorem}

The theorem above is slightly different from the original statement in \cite{BuzekHilleryWerner1999}. Bu\v{z}ek, Hillery, and Werner state the bound for \emph{universal}
channels, whose fidelity is \[\Tr[(I - P)\Lambda(P^{\otimes N})]\] and it is the same
for every $P$. However, this issue is easy to fix. The bound in Theorem \ref{thm:fm-unot} (the averaged form) follows by the following argument: for an arbitrary
channel $\Lambda$, the twirled channel
\[\widetilde\Lambda(X) = \int_{\mathrm{SU}(2)} V^\dagger \Lambda\bigl(V^{\otimes N} X\, V^{\dagger \otimes N}\bigr) V\, dV\]
is universal, and by the invariance of the measure, its fidelity at every $P$
equals the average fidelity of $\Lambda$. The transposition and the universal-NOT differ only by a fixed unitary. By
\eqref{eq:fm-unot-transpose} and $Y^2 = I$,
\begin{equation}\label{eq:fm-transpose-unot}
  P^\top = Y (I - P)\, Y \qquad \text{for every pure state } P .
\end{equation}
Hence, post-composing a transposition with conjugation by $Y$ turns it into a
universal-NOT with the same fidelity, and Theorem~\ref{thm:fm-unot}
transfers directly to transposition.

\begin{figure}[t]
\centering
\includegraphics[width=0.9\linewidth]{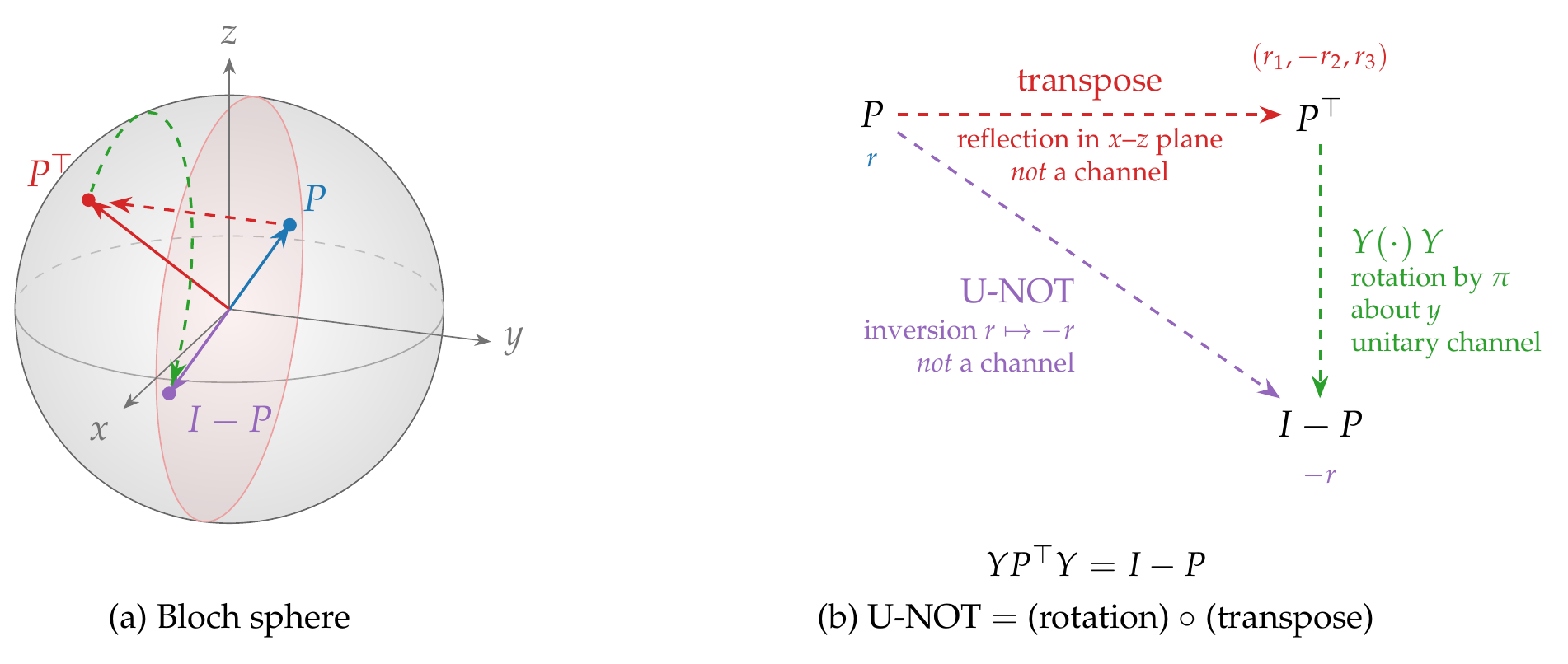}
\caption{Transposition and the universal-NOT differ by a fixed unitary.
Solid arrows are Bloch vectors and dashed arrows are maps.
(a)~For a pure state $P$ with Bloch vector $r$ (blue), transposition reflects
$r$ through the $x$--$z$ plane (shaded; dashed red arrow) to $(r_1,-r_2,r_3)$, the Bloch vector
of $P^\top$ (red). Conjugation by the Pauli matrix $Y$ rotates the sphere by
$\pi$ about the $y$-axis (dashed green arc) and sends this vector to $-r$, the Bloch
vector of the orthogonal state $I - P$ (purple, behind the sphere).
(b)~The two orientation-reversing maps, transposition and the U-NOT
$r \mapsto -r$, are not channels, but they differ by the unitary channel
$Y(\cdot)\,Y$, so $Y P^\top Y = I - P$. Hence any channel that approximates
one of them yields a channel that approximates the other with the same
fidelity.}
\label{fig:fm-unot}
\end{figure}

\begin{corollary}[Transposition from finitely many copies]
\label{cor:fm-transpose}
For every $N \ge 1$ and every channel
$\Lambda : \Lin\bigl((\mathbb{C}^2)^{\otimes N}\bigr) \to \Lin(\mathbb{C}^2)$,
\[
  \int \Tr\bigl[P^\top \Lambda(P^{\otimes N})\bigr]\, dP \le \frac{N + 1}{N + 2},
\]
equivalently
\[
  \int \Bigl(1 - \Tr\bigl[P^\top \Lambda(P^{\otimes N})\bigr]\Bigr)\, dP
  \ge \frac{1}{N + 2}.
\]
\end{corollary}

\begin{proof}
The map $\Lambda_Y(X) = Y \Lambda(X)\, Y$ is a channel, since it is
$\Lambda$ followed by a unitary conjugation. By~\eqref{eq:fm-transpose-unot}
and cyclicity of the trace,
\[
  \Tr\bigl[P^\top \Lambda(P^{\otimes N})\bigr]
  = \Tr\bigl[Y (I - P) Y\, \Lambda(P^{\otimes N})\bigr]
  = \Tr\bigl[(I - P)\, \Lambda_Y(P^{\otimes N})\bigr].
\]
Integrating over $P$ and applying Theorem~\ref{thm:fm-unot} to
$\Lambda_Y$ gives the claim.
\end{proof}

The same argument run backwards shows that the two problems are
equivalent, so Corollary~\ref{cor:fm-transpose} is tight: the optimal
transposer estimates $P$ and prepares the transpose of the estimate. This
is the ``identify, then prepare'' principle of
Section~\ref{sec:finite-classical}. For a continuum of states,
identification is never exact, and the error decays only like $1/N$; see
also Buscemi et al.~\cite{Buscemi2003} for optimal approximations of the
transpose map.

\subsection{From Nash convergence to fast transposition}\label{sec:fm-combine}

We now prove Theorem~\ref{thm:fm-barrier}. The proof is by contradiction.
Assuming that Alice's learner converges against every pure state, we use
it as a subroutine to build, for each $m \ge 1$, a channel $\Lambda_m$ that
transposes $N = (b + 3)m$ copies of an \emph{arbitrary} pure state with
an error that is exponentially small in $m$. This contradicts
Corollary~\ref{cor:fm-transpose}, whose bound decays only like $1/m$. The
channel $\Lambda_m$ is a device used in the proof and not part of the model. We first explain why such a device is needed and how
it is built, and then give the proof.

\paragraph{Two gaps between the learner and the transposition bound.}
Suppose that $f_t(P) \to 0$ for every pure state $P$. We then have two
facts that pull in opposite directions.
\begin{itemize}
  \item By Lemma~\ref{lem:fm-patch}, there is an open patch $U$ of pure
        states and constants $C < \infty$ and $q < 1$ such that the
        learner's error after $m$ rounds satisfies $f_m(P) \le C q^m$ for
        every $P \in U$.
  \item By Corollary~\ref{cor:fm-transpose}, every channel that maps $N$
        copies of an unknown pure state to one qubit has transposition
        error at least $1/(N + 2)$, on average over \emph{all} pure states.
\end{itemize}
These facts cannot be compared for two reasons. First, the corollary
concerns a single channel acting on $N$ copies, whereas the learner is a
process that runs over many rounds. Second, the corollary averages over
the whole Bloch sphere, whereas the learner is known to be fast only on
the patch $U$, which may be very small. Outside $U$ we know only that
$f_m \le 1$, and this is useless inside an average.

The channel $\Lambda_m$ closes both gaps. The first gap is closed by
bookkeeping: the $m$ rounds of the learner against a fixed state $P$ consume
$bm$ copies of $P$ and produce the state $\Sigma_m(P)$, so ``run the
learner for $m$ rounds and read out the strategy'' is itself a channel
from mapping $bm$ qubits to one qubit. The second gap is closed by a symmetry of
transposition, which lets us move any input state into the patch before
the learner sees it, and move the answer back afterwards.

\paragraph{Rotations and transposition.}
Write $\vec\sigma = (X, Y, Z)$ for the Pauli matrices and, for $r \in S^2$,
let $P(r) = \frac12\bigl(I + r \cdot \vec\sigma\bigr)$ be the pure state
with Bloch vector $r$. Every unitary $V$ on $\mathbb{C}^2$ acts on Bloch
vectors by a rotation $O_V \in \mathrm{SO}(3)$, that is,
$V P(r) V^\dagger = P(O_V r)$, and every rotation arises in this way.

Transposition is compatible with rotations: if the input is rotated, the
output is rotated as well, by a rotation that we know. Precisely, since
the transpose reverses the order of a product,
\[
  (V P V^\dagger)^\top = \overline{V}\, P^\top V^\top ,
\]
so rotating the input by $V$ rotates its transpose by $\overline{V}$.
Because $V^\top$ is unitary with inverse $\overline{V}$, we can solve for
$P^\top$:
\begin{equation}\label{eq:fm-rotate-back}
  P^\top = V^\top\, (V P V^\dagger)^\top\, \overline{V}
  \qquad \text{for every pure state } P \text{ and every unitary } V .
\end{equation}
In words, to transpose $P$ one may first rotate it by $V$, then transpose
the rotated state, and finally rotate back by conjugating with $V^\top$.
The same recipe works for approximate transposition, with no loss of
fidelity: for every pure state $P$, unitary $V$, and state $\tau$,
\begin{equation}\label{eq:fm-covariance}
  \Tr\bigl[P^\top\, V^\top \tau\, \overline{V}\bigr]
  = \Tr\bigl[(V P V^\dagger)^\top\, \tau\bigr],
\end{equation}
which follows from~\eqref{eq:fm-rotate-back} and the cyclicity of the
trace. Thus, if $\tau$ is a good approximation of the transpose of the
rotated state $V P V^\dagger$, then $V^\top \tau\, \overline{V}$ is an
equally good approximation of $P^\top$. Hence a device that transposes
well on the patch $U$ also transposes well at any other state $P$,
{given that we know a rotation that carries $P$ into $U$}.

\paragraph{The construction of $\Lambda_m$.}
The input state is unknown, so the rotation must be chosen from the input
itself. This costs some copies, but only a rough estimate is needed: the
patch has a fixed size, and it suffices to locate the Bloch vector of the
input up to a fixed accuracy. We prepare this choice in advance. Since
$U$ is open, there are $r^* \in S^2$ and $\delta \in (0, \tfrac14)$ such
that
\begin{equation}\label{eq:fm-patch-ball}
  P(r) \in U \qquad \text{whenever } \|r - r^*\|_2 < 3\delta .
\end{equation}
By compactness, $S^2$ has a finite $\delta$-net $\{v_1, \dots, v_K\}$,
that is, every point of $S^2$ lies within distance $\delta$ of some $v_k$.
For each $k$ we fix a unitary $V_k$ whose rotation carries $v_k$ to the
center of the patch, $O_{V_k} v_k = r^*$. The numbers $\delta$ and $K$,
the net, and the unitaries depend only on $U$, and not on $m$ or on the
input.

Fix $m \ge 1$ and let $N = (b + 3)m$. The channel
$\Lambda_m : \Lin\bigl((\mathbb{C}^2)^{\otimes N}\bigr) \to \Lin(\mathbb{C}^2)$
splits its $N$ input qubits into $3m$ qubits for estimation and $bm$
qubits for the learner, and it acts in four stages.
\begin{enumerate}
  \item \textbf{Estimate.} Measure $X$ on $m$ of the estimation qubits, $Y$
        on $m$ more, and $Z$ on the last $m$, and let
        $y = (y_X, y_Y, y_Z) \in [-1, 1]^3$ be the vector of empirical
        means of the $\pm 1$ outcomes. On the input $P(r)^{\otimes N}$, the
        vector $y$ is an estimate of $r$.
  \item \textbf{Choose a rotation.} Let $k = \mathrm{ind}(y) \in \{1, \dots, K\}$
        be the index of a net point nearest to $y/\|y\|_2$, with a fixed
        tie-breaking rule and $\mathrm{ind}(0) = 1$. If the estimate is
        accurate, then $v_k$ is close to $r$, so $V_k$ carries the input
        state close to the center of the patch.
  \item \textbf{Rotate and learn.} Apply $V_k$ to each of the remaining
        $bm$ qubits. Then run Alice's learner for $m$ rounds, starting
        from $\Psi_0$ and feeding it $b$ of the rotated qubits per round,
        and apply the readout channel $R$. This stage is the map
        $X \mapsto \mathcal{A}(X \otimes \,\cdot\,)$ applied $m$ times,
        followed by $R$.
  \item \textbf{Rotate back.} Conjugate the resulting qubit by $V_k^\top$,
        that is, apply $\tau \mapsto V_k^\top \tau\, \overline{V_k}$.
\end{enumerate}
Each stage is a measurement, a unitary, or a channel, possibly controlled
by the classical outcome $k$, so $\Lambda_m$ is a channel. Stages~1 and~2
find a rotation that carries the unknown input into the patch. Stage~3
runs the learner there, where it is fast. Stage~4 converts the learner's
answer for the rotated state into an answer for the original state, as
in~\eqref{eq:fm-covariance}. The device has two sources of error: the
learner's error on the patch, and the event that the estimate is so poor
that the rotated state misses the patch. The proof shows that both are
exponentially small in $m$.

\begin{proof}[Proof of Theorem~\ref{thm:fm-barrier}]
Suppose, for contradiction, that $f_t(P) \to 0$ for every pure state $P$.
Let $U$, $C$, and $q$ be as in Lemma~\ref{lem:fm-patch}, and let
$\Lambda_m$ be the channel constructed above.\vspace{0.5em}

\noindent\textbf{Step 1: the error of $\Lambda_m$ is an average of
learner errors.} Fix a pure state $P$ and consider the input
$P^{\otimes N}$. This is a product state, so the outcome $k$ of Stages~1
and~2 is independent of the $bm$ qubits that were not measured, and these
qubits remain in the state $P^{\otimes bm}$. Let
$p_k(P) = \Pr[\mathrm{ind}(y) = k]$. On outcome $k$, Stage~3 feeds the
learner $b$ copies of the rotated state $V_k P V_k^\dagger$ in each of $m$
rounds. By~\eqref{eq:fm-FP}, the learner cannot distinguish this from
$m$ rounds of play against an opponent with the fixed strategy
$V_k P V_k^\dagger$, so its readout is exactly
$\Sigma_m(V_k P V_k^\dagger)$. After Stage~4,
\[
  \Lambda_m(P^{\otimes N})
  = \sum_{k=1}^K p_k(P)\; V_k^\top\, \Sigma_m\bigl(V_k P V_k^\dagger\bigr)\, \overline{V_k}.
\]
By~\eqref{eq:fm-covariance} with $\tau = \Sigma_m(V_k P V_k^\dagger)$, the
fidelity of the $k$-th term with $P^\top$ equals the fidelity of
$\Sigma_m(V_k P V_k^\dagger)$ with $(V_k P V_k^\dagger)^\top$, which is
$1 - f_m(V_k P V_k^\dagger)$ by~\eqref{eq:fm-error}. Hence the infidelity
of the device is
\begin{equation}\label{eq:fm-device-error}
  \varepsilon_m(P) \coloneqq 1 - \Tr\bigl[P^\top \Lambda_m(P^{\otimes N})\bigr]
  = \sum_{k=1}^K p_k(P)\, f_m\bigl(V_k P V_k^\dagger\bigr).
\end{equation}
The device errs exactly as much as the learner does at the rotated state,
averaged over the choice of rotation.

\noindent\textbf{Step 2: a good rotation gives an exponentially small
error.} Let $P = P(r)$, and call an index $k$ \emph{good for $r$} if
$\|r - v_k\|_2 \le 2\delta$. For a good $k$, the rotated state is
$V_k P V_k^\dagger = P(O_{V_k} r)$. Rotations preserve distances and
$O_{V_k} v_k = r^*$, so
\[
  \|O_{V_k} r - r^*\|_2 = \|O_{V_k}(r - v_k)\|_2 = \|r - v_k\|_2 \le 2\delta .
\]
By~\eqref{eq:fm-patch-ball}, the rotated state lies in $U$, and therefore
$f_m(V_k P V_k^\dagger) \le C q^m$. For the other indices we use only
$f_m \le 1$. Splitting the sum in~\eqref{eq:fm-device-error} into good and
bad indices gives
\begin{equation}\label{eq:fm-device-split}
  \varepsilon_m(P) \le C q^m + \Pr\bigl[\mathrm{ind}(y) \text{ is not good for } r\bigr].
\end{equation}

\noindent\textbf{Step 3: the rotation is good with overwhelming
probability.} Each component of $y$ is the mean of $m$ independent
$\pm 1$ outcomes whose expectation is the corresponding component of $r$,
since $\Tr[A\, P(r)]$ is the $X$-, $Y$-, or $Z$-component of $r$ for
$A = X$, $Y$, or $Z$. Hoeffding's inequality~\cite{Hoeffding1963} bounds
the probability that one component deviates by more than $a$ by
$2e^{-ma^2/2}$. Taking $a = \delta/(2\sqrt3)$ and a union bound over the
three components,
\begin{equation}\label{eq:fm-hoeffding}
  \Pr\Bigl[\|y - r\|_\infty > \tfrac{\delta}{2\sqrt3}\Bigr]
  \le 6\, e^{-m\delta^2/24}.
\end{equation}
On the complementary event, $\|y - r\|_2 \le \sqrt3\,\|y - r\|_\infty \le \delta/2 < 1$,
so $y \neq 0$, and since $\|r\|_2 = 1$,
\[
  \Bigl\| \tfrac{y}{\|y\|_2} - r \Bigr\|_2
  \le \bigl|\, 1 - \|y\|_2 \bigr| + \|y - r\|_2
  \le 2\|y - r\|_2 \le \delta .
\]
The selected net point lies within $\delta$ of $y/\|y\|_2$, because the
$v_k$ form a $\delta$-net. It therefore lies within $2\delta$ of $r$, so
$\mathrm{ind}(y)$ is good for $r$. Thus a bad index occurs with
probability at most $6e^{-m\delta^2/24}$, and~\eqref{eq:fm-device-split}
becomes
\begin{equation}\label{eq:fm-device}
  \varepsilon_m(P) \le C q^m + 6\, e^{-m\delta^2/24}
  \qquad \text{for every pure state } P \text{ and every } m \ge 1 .
\end{equation}
The right-hand side does not depend on $P$: the device transposes every
pure state, and not only the states of the patch, with exponentially small
error.

\noindent\textbf{Step 4: the contradiction.} The channel $\Lambda_m$ acts
on $N = (b + 3)m$ copies, so Corollary~\ref{cor:fm-transpose} bounds its
average error from below. Averaging~\eqref{eq:fm-device} over $P$,
\[
  \frac{1}{(b + 3)m + 2}
  \;\le\; \int \varepsilon_m(P)\, dP
  \;\le\; C q^m + 6\, e^{-m\delta^2/24}
  \qquad \text{for every } m \ge 1 .
\]
Multiplying by $m$ and letting $m \to \infty$, the left-hand side tends to
$1/(b + 3) > 0$, while the right-hand side tends to $0$. This
contradiction shows that $f_t(P) \not\to 0$ for some pure state $P$.
\end{proof}



\paragraph{Consequence for uncoupled game dynamics.}
Consider the continuum of conjugation games $G_P$, indexed by pure qubit
states $P$, with $R_A = 2\ket{\Omega}\bra{\Omega}$ and $R_B^P = I \otimes P$.
By Lemma~\ref{lem:prelim-conj-ne}, $P$ is Bob's unique dominant strategy,
and $G_P$ has the unique Nash equilibrium $(P^\top, P)$.
Theorem~\ref{thm:fm-barrier} says directly that no finite-memory learner
for Alice converges to her best response against every stationary opponent
in this family. To obtain a statement about Nash convergence when both
players learn, we need Bob to remain stationary once he plays his dominant
strategy.

\begin{definition}[Retention of a dominant strategy]
A rule \emph{retains a strictly dominant strategy} if, whenever the player's
current strategy is their unique dominant strategy, the next strategy is
the same, for every state of the player's private memory and every
opponent profile.
\end{definition}

\begin{corollary}[Finite-memory barrier for uncoupled dynamics]
\label{cor:fm-games}
Consider autonomous uncoupled dynamics for the games $\{G_P\}$, in which
Alice has a finite-dimensional persistent register and receives $b$ copies
of Bob's current strategy per round, with $b \ge 1$ fixed. If Bob's rule
retains his strictly dominant strategy, then the dynamics do not converge
to Nash equilibrium from every initial strategy profile in every game of
the family.
\end{corollary}

\begin{proof}
For each game $G_P$, start Bob at $P$. By retention, Bob plays $P$ in every
round, whatever his private memory does. Alice's payoff $R_A$ is the same
in every game, so uncoupledness gives her the same channel, readout, and
initial state throughout the family. Her trajectory is therefore exactly
$\Sigma_t(P) = R(F_P^{\,t}(\Psi_0))$. Convergence to the equilibrium in
every game would require $\Sigma_t(P) \to P^\top$ for every pure $P$, which
contradicts Theorem~\ref{thm:fm-barrier}.
\end{proof}

\begin{remark}[Retention cannot be dropped]\label{rem:fm-retention}
Bob knows $P$ classically from his own payoff, so he can prepare $P^\top$.
Without retention, he could play $P^\top$ in the first round, which hands
Alice her answer, and switch to $P$ afterwards, while Alice copies Bob's
first state into her strategy register and then keeps it. With a two-label
phase register for Bob and a three-label phase register for Alice, this
pair of rules is autonomous and uncoupled, and it reaches $(P^\top, P)$
within two rounds in every game. Retention excludes exactly this kind of
signaling through Bob's own strategy.
\end{remark}

\section{Conclusion and Future Directions}
\label{sec:conclusion}

We studied uncoupled learning when strategies are quantum states, and every
update must be a physical operation on copies of the opponent's state. As
a warm-up, we showed that the classical obstruction of Hart and Mas-Colell
carries over to qubit strategies when the players have classical access to
each other's states (Theorem~\ref{thm:warmup}). Our main results concern
conjugation games, whose classical counterparts are trivial, so that the
only difficulty is physical realizability. In these games, the possibility
of Nash convergence is governed by Alice's memory. Without memory, the
convergence already fails in four single-qubit games, and any channel that
keeps the four equilibria fixed leaves Alice's strategy frozen
(Theorems~\ref{thm:four-game} and~\ref{thm:nqubit}). With $O(\log L)$
classical bits, a guess-and-test rule converges in every family of $L$
pure targets, with an explicit mixing time (Theorem~\ref{thm:fc}). With
any finite memory, classical or quantum, convergence fails for the
continuum of all pure qubit targets (Theorem~\ref{thm:fm-barrier} and
Corollary~\ref{cor:fm-games}). All of these barriers concern exact
convergence, and the case of approximate equilibria remains open.

The proofs rest on two tools. Stinespring's dilation records what a
channel discards into its environment, and the decomposition of
power-bounded operators into a recurrent and a transient part forces a
finite-memory learner to converge exponentially fast or not at all. We
close with two directions in which these tools may be useful beyond our setting.

\paragraph{Stinespring's dilation and the forgetfulness of learning
dynamics.}
A recent line of work relates the speed of last-iterate convergence to how
quickly an algorithm forgets its past. Optimistic gradient
descent-ascent (OGDA) and optimistic multiplicative weights update (OMWU)
both converge in the last iterate in zero-sum
games~\cite{DaskalakisPanageas2019,WeiEtAl2021}, and OGDA does so at a
rate that is uniform over all games of a given
size~\cite{CaiOikonomouZheng2022}. Cai et al.~\cite{CaiEtAl2024} show that
no such uniform rate exists for OMWU, nor for other optimistic
follow-the-regularized-leader algorithms that do not forget the past
quickly: fast last-iterate convergence requires a \emph{forgetful}
algorithm. For quantum players, remembering and forgetting can be
accounted for exactly. By Theorem~\ref{thm:prelim-stinespring}, an update
channel embeds its input isometrically into the learner's persistent
register and an environment. What reaches the register is remembered, and
what reaches the environment is forgotten. 

Our results contain two
extreme cases. A memoryless learner that preserves the equilibria
deposits an isometric copy of Bob's state in the environment and so
forgets it entirely (Lemma~\ref{lem:four-state}), whereas the
guess-and-test rule of Section~\ref{sec:finite-classical} retains a single
label and discards everything else. Moreover, our learners are quantum
channels with memory in the sense of Kretschmann and
Werner~\cite{KretschmannWerner2005}, who call such a channel
\emph{forgetful} when the effect of its initial memory dies out over time.
This suggests the following questions. Can the forgetfulness
of~\cite{CaiEtAl2024} be expressed through the dilation, for instance by
the information that a learner releases to its environment in each
round~\cite{KretschmannSchlingemannWerner2008}, and is there a
quantitative trade-off between this information and the rate of
convergence? Which learning algorithms for quantum zero-sum games, such
as matrix multiplicative weights and its optimistic
variants~\cite{JainWatrous2009,VasconcelosEtAl2025}, are forgetful in this
sense? Our results add a constraint in the opposite direction, since a
learner that forgets too much cannot converge at all. A quantum learner
must therefore remember enough to identify the equilibrium and forget
enough to reach it quickly.

\paragraph{Power-bounded operators and the convergence of learning
dynamics.}
Lemma~\ref{lem:fm-exp} is a statement about any power-bounded linear map:
a linear observable of its orbit tends to zero exponentially fast or not
at all. It applies to Alice's learner because, against a stationary
opponent, her register evolves by powers of a single channel. Standard
learning dynamics have the same structure in the ambient space. Consider a
bilinear zero-sum game with payoff $x^\top A y$, and let $z = (x, y)$ and
$J = \bigl(\begin{smallmatrix} 0 & A \\ -A^\top & 0 \end{smallmatrix}\bigr)$.
Without constraints, OGDA is the recursion
\begin{equation}\label{eq:conclusion-ogda}
  z_{t+1} = z_t - \eta\, J\,(2 z_t - z_{t-1}),
\end{equation}
which is a single linear map on the pair $(z_t, z_{t-1})$. For
$\eta \|J\| < \tfrac12$ this map is power-bounded. Its recurrent part
consists of the equilibria, which are fixed points, and its transient part
decays geometrically. The decomposition of Lemma~\ref{lem:fm-split} thus
recovers the linear last-iterate convergence of unconstrained
OGDA~\cite{DaskalakisEtAl2018,LiangStokes2019}. OMWU follows the same
recursion in the dual space of cumulative payoffs, with $z_t$ obtained
from the dual variable by the softmax map. On the simplex, therefore,
neither algorithm is linear: OGDA composes the linear step with a
Euclidean projection, and OMWU composes it with the softmax map. The same
happens in our model as soon as both players learn, because the
dynamics~\eqref{eq:prelim-memoryless} are linear on
$\Lin(H_A \otimes H_B)$ but are applied only to product states
$\rho_t \otimes \sigma_t$, on which they are quadratic. This is one reason
why our finite-memory barrier is proved against a stationary opponent.

Once linearity is lost, Lemma~\ref{lem:fm-exp} no longer applies, and the
picture is richer. Regularized learning is recurrent in zero-sum games,
both classical~\cite{MertikopoulosPapadimitriouPiliouras2018} and
quantum~\cite{JainPiliourasSim2022}, OGDA converges at a rate that is
uniform over games, and OMWU converges without such a
rate~\cite{CaiEtAl2024}. The distinction between a rate for each game and
a rate that is uniform over a family is also the crux of
Section~\ref{sec:barrier}, where passing from the former to the latter
required the Baire category theorem (Lemma~\ref{lem:fm-patch}). It is
natural to ask whether a recurrent/transient decomposition still organizes
these behaviors. One route is to restore linearity by passing from the
update map to its Koopman operator, which acts linearly and contractively
on functions of the state and is therefore power-bounded. The
Jacobs--de~Leeuw--Glicksberg decomposition is available for such operators
on suitable function spaces~\cite{EisnerFarkasHaaseNagel2015}, but in
infinite dimensions, its stable part carries no rate. We ask whether
recurrence corresponds to the reversible part of the decomposition and
whether the gap between forgetful and non-forgetful algorithms corresponds
to the presence (or absence) of a uniform spectral gap on its stable part. A
positive answer would extend the principle ``exponentially fast or not at
all'' from finite-memory quantum learners to a classification of
convergence behavior for general learning dynamics.
\bibliographystyle{abbrv}
\bibliography{ref}
\end{document}